\documentclass[useAMS,usenatbib]{mn2e}
\usepackage{epsfig}
\usepackage{amsfonts}
\usepackage[usenames,dvipsnames]{xcolor}
\usepackage{url}
\usepackage{subfigure}

\usepackage{times,graphicx,amsmath,amsfonts,amssymb,aas_macros,epstopdf}
\newcommand{\mpl}{M_{\rm Pl}}
\newcommand{\rd}{{\rm d}}

\def\eg{{\frenchspacing\it e.g.}}

\def\be{\begin{equation}}
\def\ee{\end{equation}}
\def\ba{\begin{eqnarray}}
\def\ea{\end{eqnarray}}

\def\hmpc{h^{-1}\,{\rm Mpc}}
\def\hkpc{h^{-1}\,{\rm kpc}}

\newcommand{\BLED}[1]{\textcolor{black}{#1}}

\newcommand{\GZ}[1]{\textcolor{black}{#1}}
\newcommand{\beq}{\begin{equation}}
\newcommand{\eeq}{\end{equation}}

\title[Power spectra in $f(R)$ gravity]
  {The nonlinear matter and velocity power spectra in $f(R)$ gravity}
\author[Baojiu~Li~{\it et~al.}]
{\parbox[t]{15.4cm}
  {Baojiu Li$^{1}$, 
   Wojciech A.~Hellwing$^{1,2,3}$, 
   Kazuya Koyama$^{4}$,
   Gong-Bo Zhao$^{4,5}$, 
   Elise Jennings$^{6,7}$, 
   Carlton M.~Baugh$^{1}$}
   \vspace*{8pt}\ \\
  $^{1}$Institute of Computational Cosmology, Department of Physics, Durham University, South Road, Durham DH1 3LE, UK\\
  $^{2}$Interdisciplinary Centre for Mathematical and Computational Modeling (ICM), University of Warsaw, ul. Pawi\'nskiego 5a, Warsaw, Poland\\
  $^{3}$Institute of Astronomy, University of Zielona G\'ora, ul. Lubuska 2, Zielona G\'ora, Poland\\
  $^{4}$Institute of Cosmology \& Gravitation, University of Portsmouth, Portsmouth PO1 3FX, UK\\
  $^{5}$National Astronomy Observatories, Chinese Academy of Science, Beijing, 100012, P.R.China\\
  $^{6}$The Kavli Institute for Cosmological Physics, University of Chicago, 5640 South Ellis Avenue, Chicago, IL 60637, US\\
  $^{7}$The Enrico Fermi Institute, University of Chicago, 5640 South Ellis Avenue, Chicago, IL 60637, US
}

\pagerange{\pageref{firstpage}--\pageref{lastpage}} \pubyear{2012}

\def\LaTeX{L\kern-.36em\raise.3ex\hbox{a}\kern-.15em
    T\kern-.1667em\lower.7ex\hbox{E}\kern-.125emX}

\begin{document}

\label{firstpage}

\maketitle

\begin{abstract}
We study the matter and velocity divergence power spectra in a $f(R)$ gravity theory and their time evolution measured from several large-volume $N$-body simulations with varying box sizes and resolution. We find that accurate prediction of the matter power spectrum in $f(R)$ gravity places stronger requirements on the simulation than is the case with $\Lambda$CDM, because of the nonlinear nature of the fifth force. Linear perturbation theory is shown to be a poor approximation for the $f(R)$ models, except when the chameleon effect is very weak. We show that the relative differences from the fiducial $\Lambda$CDM model are much more pronounced in the nonlinear tail of the velocity divergence power spectrum than in the matter power spectrum, which suggests that future surveys which target the collection of peculiar velocity data will open new opportunities to constrain modified gravity theories. A close investigation of the time evolution of the power spectra shows that there is a pattern in the evolution history, which can be explained by the properties of the chameleon-type fifth force in $f(R)$ gravity. Varying the model parameter $|f_{R0}|$, which quantifies the strength of the departure from standard gravity, mainly varies the epoch marking the onset of the fifth force, as a result of which the different $f(R)$ models are in different stages of the same evolutionary path at any given time.
\end{abstract}


\section{Introduction}

\label{sect:intro}

The origin of the observed accelerated expansion of the Universe \citep{retal,petal} is one of the most challenging questions in contemporary theoretical physics. Although the standard cold dark matter (CDM) model plus a cosmological constant can explain this observation very well, the so-called $\Lambda$CDM paradigm suffers from serious theoretical problems, as the vacuum energy density predicted by particle physics theory is many orders of magnitude larger than the cosmologically inferred value of the cosmological constant. This has motivated the proposal of alternative models to explain the accelerated expansion of the Universe. 

So far, most of these models can be divided into two classes: dark energy \citep[see][for a review]{cst2006}, which involves one or more dynamical fields or new matter species that accelerate the expansion of the Universe, and modified gravity \citep{cfps2012}, which proposes that general relativity (GR) breaks down on cosmological scales and must be accompanied by certain modifications. Other models, such as the inhomogeneous universe model \citep{bn2008}, have also been studied as alternatives to $\Lambda$CDM, but to a lesser extent.

Unlike pressure-less matter, usually dark energy does not cluster strongly \citep[which is the case for, e.g., the quintessence model of][]{wcos2000} and its effects are mainly to modify the cosmic expansion history \citep[there are, however, exceptions, such as the coupled quintessence model of][in which the dark energy field does experience strong clustering]{a2000}. In these models, structure formation is different from {that in $\Lambda$CDM only because the background expansion rate has been modified. In contrast, modified gravity models often predict a different force law between matter particles, therefore changing structure formation directly. One can therefore in principle distinguish between these possibilities using a combination of observables \citep{jz2008}.

Any modification to the force law in modified gravity theories is highly constrained, because GR has been confirmed to high accuracy by local tests \citep[see. e.g.,][]{gr_test2,gr_test1,gr_test4,gr_test3,w2006}. If we consider the modification to the standard gravity as a new force, the so-called fifth force, then the fifth force must either have a very weak strength or very short (sub-millimeter) range, in order to be consistent with local tests. Consequently, any viable modified gravity theory must have some mechanism to suppress (or screen) the fifth force at least in regions where local tests have been carried out. In the case that the new force is mediated by a scalar degree of freedom, there are several elegant examples of such screening mechanisms, including the chameleon \citep{kw2004,ms2007}, dilaton \citep{bbds2010}, symmetron \citep{hk2010} and Veinshtein \citep{dgp2000,nrt2009,dev2009}. 

In this work, we focus on one of the most well-studied modified gravity models, $f(R)$ gravity \citep{cddett2005}, which employs the chameleon mechanism to suppress the fifth force in high-density regions. In particular, to study the behaviour of the matter and velocity divergence power spectra in $f(R)$ gravity, we perform a number of $N$-body simulations for the $f(R)$ model proposed in \citet{hs2007} with various model parameters and simulation box sizes \citep[see, e.g.,][for some other viable $f(R)$ cosmological models studied in the literature.]{s2007,lb2007,ab2007}. \citet{jblkz2012} used some of these simulations to study the form of redshift space distortions in $f(R)$ models. Here we study the power spectra behind these distortions in more detail.

The nonlinear matter power spectrum in this particular model has been studied previously by \citet{olh2008} and \citet{zlk2011}. Our study differs from these in several aspects. Firstly, the simulations used here have the largest volume up to date for models of this type and span a wide range of box sizes with good agreement between the results of each simulation. Secondly, we present the measurements of the velocity divergence power spectrum. Thirdly, we study the time evolution patterns for both the matter and velocity divergence power spectra, and relate them to the property of the chameleon fifth force and the formation of structure in hierarchical cosmologies. 

The outline of the present work is as follows: in \S~\ref{sect:fr_gravity} we briefly describe the general $f(R)$ gravity models and explain how the chameleon mechanism works. In \S~\ref{sect:sim} we give a short description of the simulation code and the technical specifications of the simulations. \S~\ref{sect:pk} contains the main results of this paper and 
% \WHED{\sout{in which \S~\ref{subsect:patterns} presents some visualisation of the density and velocity divergence fields which could help understand the results later, \S~\ref{subsect:pk_meas} explains how the power spectra have been measured \S~\ref{subsect:pk_res} discusses a new resolution issue which arises when solving the fifth force in the $f(R)$ gravity model, \S~\ref{subsect:pk_evol} shows how the matter and velocity divergence power spectra evolve in time and observes a pattern for the evolution and \S~\ref{subsect:pk_theory} gives a theoretical explanation of the observed pattern}}
we finally summarise and conclude in \S~\ref{sect:con}.

Throughout this paper we adopt the unit convention $c=1$, where $c$ is the speed of light.

\section{The $f(R)$ gravity theory}

\label{sect:fr_gravity}

This section is devoted to a brief overview of the $f(R)$ gravity theory and its properties.

\subsection{The $f(R)$ gravity model}

\label{subsect:fr}

The $f(R)$ gravity model \citep{cddett2005} is a straightforward generalisation of GR: the Ricci scalar $R$ in the Einstein-Hilbert action is replaced with an algebraic function $f(R)$\citep[see \eg,][for recent reviews]{sf2010, dt2010}:
\begin{eqnarray}\label{eq:fr_action}
S &=& \int{\rm d}^4x\sqrt{-g}\left\{\frac{\mpl^2}{2}\left[R+f(R)\right]+\mathcal{L}_m\right\},
\end{eqnarray}
in which $\mpl$ is the Planck mass, $\mpl^{-2}=8\pi G$, $G$ is Newton's constant, $g$ is the determinant of the metric $g_{\mu\nu}$ and $\mathcal{L}_m$ is the Lagrangian density for matter fields (photons, neutrinos, baryons and cold dark matter). By designing the functional form of $f(R)$ one specifies the $f(R)$ gravity model.

Varying the action Eq.~(\ref{eq:fr_action}) with respect to the metric $g_{\mu\nu}$ yields the modified Einstein equation
\begin{eqnarray}\label{eq:fr_einstein}
G_{\mu\nu} + f_RR_{\mu\nu} -\left(\frac{1}{2}f-\Box f_R\right)g_{\mu\nu}-\nabla_\mu\nabla_\nu f_R = 8\pi GT^m_{\mu\nu},
\end{eqnarray}
in which $G_{\mu\nu}\equiv R_{\mu\nu}-\frac{1}{2}g_{\mu\nu}R$ is the Einstein tensor, $f_R\equiv \rd f/\rd R$, $\nabla_{\mu}$ the covariant derivative compatible to the metric $g_{\mu\nu}$, $\Box\equiv\nabla^\alpha\nabla_\alpha$ and $T^m_{\mu\nu}$ is the energy momentum tensor for matter. One can consider Eq.~(\ref{eq:fr_einstein}) as a fourth-order differential equation, or alternatively the standard second-order equation of GR with a new dynamical degree of freedom, $f_R$, the equation of motion of which can be obtained by taking the trace of Eq.~(\ref{eq:fr_einstein})
\begin{eqnarray}
\Box f_R = \frac{1}{3}\left(R-f_RR+2f+8\pi G\rho_m\right),
\end{eqnarray}
where $\rho_m$ is the matter density. The new degree of freedom $f_R$ is sometimes dubbed {\it scalaron} in the literature \citep{zlk2011}.

Assuming that the background Universe is described by the flat Friedmann-Robertson-Walker (FRW) metric, the line element in the perturbed Universe is written as
\begin{eqnarray}
\rd s^2 = a^2(\eta)\left[(1+2\Phi)\rd\eta^2 - (1-2\Psi)\rd x^i\rd x_i\right],
\end{eqnarray}
in which $\eta$ and $x^i$ are respectively the conformal time and comoving coordinates, $\Phi(\eta,{\bf x})$ and $\Psi(\eta,{\bf x})$ are the Newtonian potential and perturbation to the spatial curvature, which are functions of both time $\eta$ and space ${\bf x}$; $a$ denotes the scale factor of the Universe and $a=1$ today.

As we are mainly interested in the large-scale structures much smaller than the Hubble scale, and since the time variation of $f_R$ is very small in the models considered below, we shall work in the quasi-static limit by neglecting the time derivatives of $f_R$. In this limit, the scalaron equation reduces to
\begin{eqnarray}\label{eq:fr_eqn_static}
\vec{\nabla}^2f_R &=& -\frac{1}{3}a^2\left[R(f_R)-\bar{R} + 8\pi G\left(\rho_m-\bar{\rho}_m\right)\right],
\end{eqnarray}
in which $\vec{\nabla}$ is the three dimensional gradient operator (to be distinguished from the $\nabla$ introduced above), and the overbar means the background value of a quantity. Note that $R$ can be expressed as a function of $f_R$.

Similarly, the Poisson equation which governs the Newtonian potential $\Phi$ can be simplified to
\begin{eqnarray}\label{eq:poisson_static}
\vec{\nabla}^2\Phi &=& \frac{16\pi G}{3}a^2\left(\rho_m-\bar{\rho}_m\right) + \frac{1}{6}a^2\left[R\left(f_R\right)-\bar{R}\right],
\end{eqnarray}
by neglecting terms involving time derivatives, and using Eq.~(\ref{eq:fr_eqn_static}) to eliminate $\vec{\nabla}^2f_R$.

According to the above equations, there are two potential effects of the scalaron on cosmology: (i) the background expansion of the Universe may be modified by the new terms in Eq.~(\ref{eq:fr_einstein}) and (ii) the relationship between gravity and the matter density field is modified, which can change the matter clustering and growth of density perturbations. Clearly, when $|f_R|\ll1$, we have $R\approx-8\pi G\rho_m$ from Eq.~(\ref{eq:fr_eqn_static}) and so Eq.~(\ref{eq:poisson_static}) reduces to the normal Poisson equation in GR; when $|f_R|$ is large, we instead have $|R-\bar{R}|\ll8\pi G|\rho_m-\bar{\rho}_m|$ and so Eq.~(\ref{eq:poisson_static}) reduces to the normal Poisson equation with $G$ rescaled by $4/3$. Note that this $4/3$ is the maximum enhancement factor of gravity in $f(R)$ models, independent of the specific functional form of $f(R)$. The choice of $f(R)$, however, is important because it governs when and on which scale the enhancement factor changes from 1 to $4/3$: scales much larger than the range of the modification to Newtonian gravity mediated by the scalaron are unaffected and gravity is not enhanced there, while on much smaller scales the $4/3$ enhancement is fully realised -- this results in a scale-dependent modification of gravity and therefore a scale-dependent growth rate of structures.

%The relationship between $\Phi$ and $\Psi$ is also changed in $f(R)$ models, with the remaining components of the modified Einstein equation giving
%\begin{eqnarray}
%\vec{\nabla}^2(\Psi-\Phi) = \vec{\nabla}^2f_R,
%\end{eqnarray}
%if one assumes that $|\bar{f}_R|\ll1$. This implies that
%\begin{eqnarray}
%\vec{\nabla}^2(\Phi+\Psi) = 8\pi G\left(\rho_m-\bar{\rho}_m\right)a^2.
%\end{eqnarray}
%Therefore the relationship between the lensing potential and the matter density perturbations remains unchanged in \GZ{such} $f(R)$ gravity models.

\subsection{The chameleon mechanism}

\label{subsect:fr_cham}

The $f(R)$ model would have been ruled out by local tests of gravity due to the factor-of-$4/3$ enhancement to the strength of Newtonian gravity. Fortunately, it is well known that, if $f(R)$ is chosen appropriately \citep{bbh2006,ftbm2007,nv2007,lb2007,hs2007,bbds2008}, the model can exploit the chameleon mechanism \citep{kw2004,ms2007} to suppress the enhancement and therefore pass the experimental constraints in high matter density regions such as our Solar system.

The essence of the chameleon mechanism is as follows: the modifications to the Newtonian gravity can be considered as an extra, or fifth force mediated by the scalaron. Because the scalaron itself is massive, the force is of the Yukawa type and is suppressed by an exponential factor $\exp(-mr)$, in which $m$ is the scalaron mass and $r$ the distance between two test masses. In high matter density environments, $m$ is very heavy and the suppression becomes very strong. In reality, this is equivalent to setting $|f_R|\ll1$ in high density regions because of the exponential suppression, which leads to the GR limit as discussed above.

As a result, the functional form of $f(R)$ is crucial to determine whether the fifth force \GZ{can be} sufficiently suppressed in high density environments. In this work we study the $f(R)$ model proposed by \citet{hs2007}, for which
\begin{eqnarray}\label{eq:hs}
f(R) = -M^2\frac{c_1\left(-R/M^2\right)^n}{c_2\left(-R/M^2\right)^n+1},
\end{eqnarray}
where $M^2\equiv8\pi G\bar{\rho}_{m0}/3=H_0^2\Omega_m$, where $H$ is the Hubble expansion rate and $\Omega_m$ is the present-day fractional density of matter. Hereafter a subscript \GZ{$_0$} always means the present day ($a=1$) value of a quantity. It was shown by \citet{hs2007} that $|f_{R0}|<0.1$ is required to evade the Solar system constraints but the exact value depends on the behaviour of $f_R$ in \GZ{galaxies} as well.

In the background cosmology, the scalaron $f_R$ always sits close to the minimum of the effective potential that governs its dynamics, defined as
\begin{eqnarray}
V_{\rm eff}\left(f_R\right) \equiv \frac{1}{3}\left(R-f_RR+2f+8\pi G\rho_m\right),
\end{eqnarray}
around which it quickly oscillates with small amplitude \citep{bdlw2012}. Therefore we have
\begin{eqnarray}
-\bar{R} \approx 8\pi G\bar{\rho}_m-2\bar{f} = 3M^2\left(a^{-3}+\frac{2c_1}{3c_2}\right).
\end{eqnarray}
To match the $\Lambda$CDM model in background evolution, we need to set
\begin{eqnarray}
\frac{c_1}{c_2} = 6\frac{\Omega_\Lambda}{\Omega_m}
\end{eqnarray}
where $\Omega_m (\Omega_\Lambda)$ is the present day fractional energy density of the dark matter (dark energy).

By taking $\Omega_\Lambda=0.76$ and $\Omega_m=0.24$\footnote{These values are used in the $f(R)$ simulations extensively in the literature, and are adopted in the simulations of this paper in order to compare with previous work.}, we find that $|\bar{R}|\approx41M^2\gg M^2$, and this simplifies the expression of the scalaron to
\begin{eqnarray}
f_R \approx -n\frac{c_1}{c_2^2}\left(\frac{M^2}{-R}\right)^{n+1}.
\end{eqnarray}
Therefore, two free parameter\GZ{s}, $n$ and $c_1/c_2^2$, completely specify the $f(R)$ model. Indeed, the latter is related to the value of the scalaron today, $f_{R0}$, as
\begin{eqnarray}
\frac{c_1}{c_2^2} = -\frac{1}{n}\left[3\left(1+4\frac{\Omega_\Lambda}{\Omega_m}\right)\right]^{n+1}f_{R0}.
\end{eqnarray}
In what follows we study three $f(R)$ models with $n=1$ and $|f_{R0}|=10^{-6}, 10^{-5}, 10^{-4}$, which will be referred to as F6, F5 and F4 respectively. These choices of $|f_{R0}|$ are meant to cover the whole parameter space that is cosmological interesting: if $|f_{R0}|>10^{-4}$ then the $f(R)$ model violates the cluster abundance constraints \citep{svh2009}, and if $|f_{R0}|<10^{-6}$ then the difference from $\Lambda$CDM would be too small to be observable in practice (see our results presented below).
\section{$N$-body simulations of $f(R)$ gravity}

\label{sect:sim}

From Eqs.~(\ref{eq:fr_eqn_static}, \ref{eq:poisson_static}) we have seen that, given the matter density field, we can solve the scalaron field $f_R$ from Eq.~(\ref{eq:fr_eqn_static}) and plug it into the modified Poisson equation (\ref{eq:poisson_static}) to solve \GZ{for} $\Phi$. Once $\Phi$ is at hand, we can difference it to calculate the (modified) gravitational force which determines how the particles move subsequently. That is exactly what we need to do in $N$-body simulations to evolve the matter distribution.

The main challenge in $N$-body simulations of models such as $f(R)$ gravity is to solve the scalaron equation (\ref{eq:fr_eqn_static}), which is in general
highly nonlinear. One way to achieve this is to use a mesh (or a set of meshes) on which $f_R$ could be solved. This implies that mesh-based $N$-body codes are most convenient. On the other hand, tree-based codes are more difficult to apply here, as we do not have an analytical formula for the modified force law (such as $r^{-2}$ in the Newtonian case) \GZ{due to the complexities stemmed from the breakdown of the superposition principal, or the invalidity of Birkhoff theorem in modified gravity.}

$N$-body simulations of $f(R)$ gravity and related theories have previously been performed by \citet{oyaizu2008,olh2008,sloh2009,zlk2011,lz2009,lz2010,schmidt2009,lb2011,bbdls2011,lh2011,dlmw2012}. However, these simulations are mostly limited by either the box size or resolution, \GZ{or both}. For this work we have run simulations using the recently developed {\tt ECOSMOG} code \citep{lztk2012}. {\tt ECOSMOG} is a modification of the mesh-based $N$-body code {\tt RAMSES} \citep{ramses}, which calculates the gravitational force by first solving the Poisson equation on meshes using a relaxation method to obtain the Newtonian potential and then differencing the potential. The code does not solve gravity by summing over the forces from nearby particles explicitly, such as tree-based codes like {\tt GADGET} \citep{Gadget1}. Additional features of the {\tt ECOSMOG} code include:

\begin{enumerate}
\item The adaptive mesh refinement (AMR), which refines a mesh cell (i.e., splits it into 8 children cells) if the number of particles in a cell exceeds a pre-defined number (the refinement criterion). This gives a higher force resolution in high matter density regions where the chameleon effect is strong and the $f(R)$ equation is more nonlinear. The refinement criterion is normally chosen as a number between 8 and 12, and in our simulations we use 9. We find that this refinement criterion works well in our case, namely, it gives the required force resolution without generating an overly large computational overhead.
\item The multigrid relaxation algorithm that ensures quick convergence. The relaxation method finds the solution to an elliptical partial differential equation (PDE) on a mesh by iteratively updating the initial guess until it converges, i.e., becomes close enough to the true solution. But the rate of converges slows down quickly after the first few iterations. To improve on this, one can coarsify the PDE, i.e., move it to a coarser mesh, solve it there and use the coarse solution to improve the solution on the original fine mesh. Unlike other codes, {\tt ECOSMOG} does this on all the AMR meshes, greatly improving the convergence behaviour \GZ{over the whole computational domain}.
\item The massive parallelisation which makes the computation very efficient. This is the key feature that enables us to run large simulations such as the ones employed in this study, which are beyond the reach of serial codes, like the ones developed by \citet{lz2009,lz2010,lb2011}.
\end{enumerate}
A convergence criterion is used to determine when the relaxation method has converged. In {\tt ECOSMOG}, convergence is considered to be achieved when the residual of the PDE, i.e., the difference between the two sides of the PDE, is smaller than a predefined parameter $\epsilon$. We have checked that for $\epsilon<10^{-8}$ the solution to the PDE no longer changes significantly when $\epsilon$ is further reduced, and our choices of $\epsilon$ will be listed in Table~\ref{table:simulations}. Further details can be found in \citet{lztk2012}.

In this work we study the matter density and velocity divergence power spectra in the $f(R)$ cosmology over a wide range of scales and redshifts using simulations. All our numerical experiments are described by the same set of cosmological parameters, i.e. the background cosmology for all models is the same. The values of cosmological parameters for our runs are
the following: $\Omega_m = 0.24$, $\Omega_\Lambda = 0.76$, $h = 0.73$, $n_s = 0.958$ and $\sigma_8 = 0.77$. The first two are the present day values of the dimensionless energy
density of the non-relativistic matter (including baryonic and dark) and dark energy, $h$ is the dimensionless Hubble parameter today, $n_s$ is
the scalar index of the primordial power spectrum and $\sigma_8$ is \GZ{the linear rms density fluctuation measured in spheres of radius 8$h^{-1}$Mpc at $z=0$.} All models in each simulation share the same initial condition ccomputed \GZ{at} the initial time of $z_i = 49$ using the Zel'dovich approximation \citep{za}. Note that in general the modified gravity affects the generation of the initial condition too \citep{lb2011}. Here we use the same initial conditions for all models within a given simulation set because
the differences in clustering between GR and all our $f(R)$ models are negligible \GZ{at} the starting redshift.

The fact that we use the same initial conditions for simulations in a given set is an advantage. Since the initial density fields for the GR and $f(R)$ simulations have the same phases, any difference in the power spectra that we find at later times will be a direct consequence of the different dynamics between the two cosmologies. We give more details describing our numerical experiments in Table \ref{table:simulations}.

\begin{table*}
\caption{Some technical details of the simulations performed for this work. F6, F5 and F4 are respectively the labels of the $f(R)$ models with $|f_{R0}|=10^{-6}, 10^{-5}, 10^{-4}$. Here $k_{Nyq}$ denotes the Nyquist frequency. $\epsilon$ is the residual for the Gauss-Seidel relaxation used in the code \citep{lztk2012}, and the two values of the convergence criterion are for the coarsest level and refinments respectively. We also list in the last column the number of realisations for each simulation.}
\begin{tabular}{@{}lcccccc}
\hline\hline
models & $L_{\rm box}$ & no. of particles & $k_{Nyq}$ $[h/\textrm{Mpc}]$ & force resolution [$\hkpc$]& convergence criterion &realisations \\
\hline
$\Lambda$CDM, F6, F5, F4 & $1.5h^{-1}$Gpc & $1024^3$ & 2.14 & 22.9 & $|\epsilon|<10^{-12}/10^{-8}$ & $6$ \\
$\Lambda$CDM, F6, F5, F4 & $1.0h^{-1}$Gpc & $1024^3$ & 3.21 & 15.26 & $|\epsilon|<10^{-12}/10^{-8}$ & $1$ \\
$\Lambda$CDM, F6, F5, F4 & $500h^{-1}$Mpc & $512^3$ & 3.21 &  30.52 & $|\epsilon|<10^{-12}/10^{-8}$ & $1$ \\
$\Lambda$CDM, F6, F5, F4 & $250h^{-1}$Mpc & $512^3$ & 6.43 & 7.63 & $|\epsilon|<10^{-12}/10^{-8}$ & $1$ \\
\hline
\end{tabular}
\label{table:simulations}
\end{table*}

%\subsection{The clustering and flow patterns}
\begin{figure*}
 \includegraphics[width=160mm]{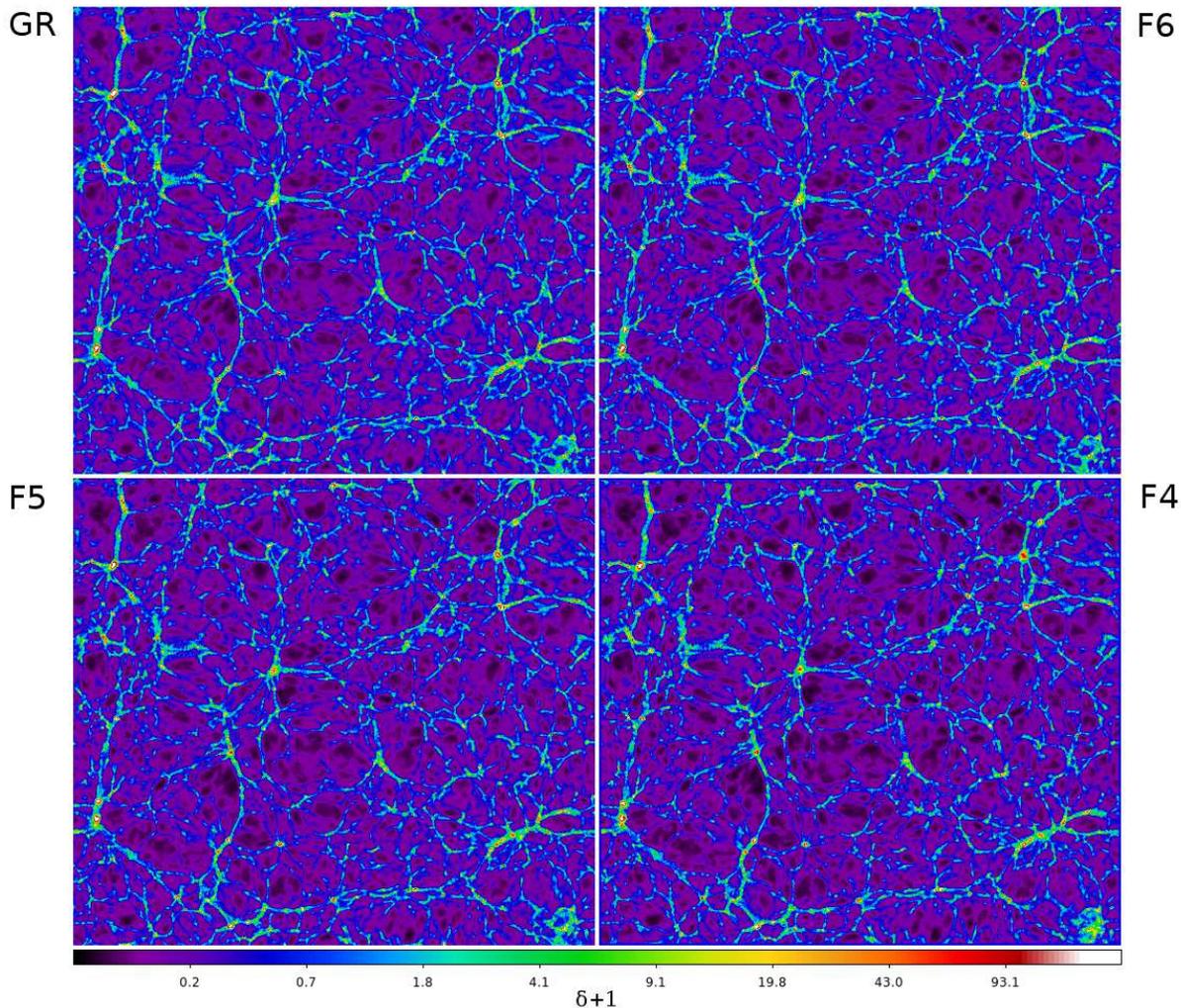}
\caption{(Colour Online) All model comparison of $z=0$ density fields ($\rho_m/\bar{\rho}_m=1+\delta$) for the $250\hmpc$ box. Each panel show a very thin slice ($\sim 0.5\hmpc$) through the DTFE density field. The top panels are results for GR (left) and F6 (right), and the bottom panels show the results for F5 (left) and F4 (right) respectively.}
\label{fig:dens_compare}
\end{figure*}
\begin{figure*}
 \includegraphics[width=160mm]{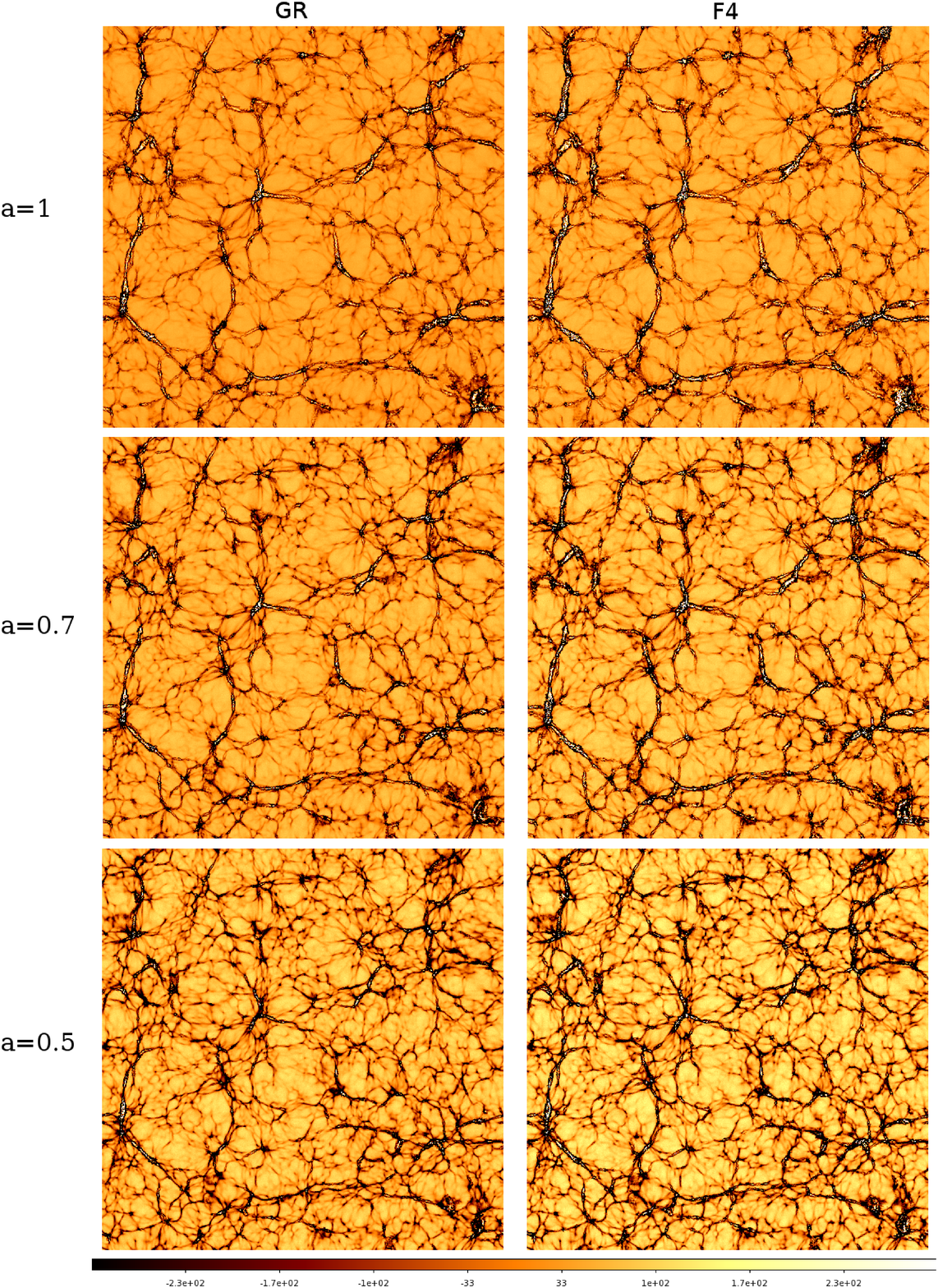}
\caption{(Colour Online) Comparison of the velocity divergence fields for the GR (left panels) and F4 (right panels) models. Each panel
shows a thin slice from the $250\hmpc$ box, and each row corresponds to a different cosmic time as labeled: $a=1$ (top), $a=0.7$ (middle) and $a=0.5$ (bottom).}
\label{fig:divv_compare}
\end{figure*}

\section{The power spectra of $f(R)$ gravity}

\label{sect:pk}

We start by introducing the dark matter density field, given by the expression
\beq
\label{eqn:deltarho}
\rho(\vec{x},t) = \left<\rho\right>\,(1 + \delta) \, ,
\eeq
where $\left<\rho(t)\right>$ is
the ensemble average of the dark matter density at time $t$, and $\delta(\vec{x},t)$ describes local deviations from homogeneity.
Structure formation is driven only by the spatially fluctuating part of the gravitational potential, $\phi(\vec{x},t)$,
induced by the density fluctuation field $\delta$. In $f(R)$ cosmologies, however, we expect that in regions where
the fifth force is not screened by the chameleon mechanism we will have an additional boost to the standard gravitational
potential induced by the scalaron as described by Eq. (\ref{eq:poisson_static}). Thus we expect that to some extent
clustering will be enhanced in our $f(R)$ models. A convenient measure of the strength of dark matter clustering is
the power spectrum. For a Fourier representation of a real space density field
\beq
\label{eqn:delK}
\delta_{\vec{k}} \, \equiv \, (2\pi)^{-3/2}\,\int \delta(\vec{x})\,e^{-i\vec{k}\cdot\vec{x}}\,d^3\vec{x}\,\,,
\eeq
the power spectrum is defined as (assuming spatial isotropy)
\beq
\label{eqn:P(k)}
P_{\delta\delta}(k) \, \equiv \, P(k) \, = \, \left< |\delta_{\vec{k}}|^2\right> \, .
\eeq
In addition to the measure of dark matter clustering we are also interested in the statistical measure of the cosmic peculiar velocity field.
The irrotational velocity field, $v(\vec{x})$, can be characterised, up to an additive bulk velocity, by a single scalar field
such as the velocity divergence
\beq
\label{eqn:vel_div}
\theta(\vec{x})={1\over H}\nabla\cdot v(\vec{x})\,.
\eeq
The $\theta$ is called the \textit{expansion scalar} \citep[see, e.g.,][]{1980Peebles}. Division of the velocity divergence
by the Hubble constant makes this quantity dimensionless. The Fourier transform of the real space expansion scalar is
\beq
\label{eqn:thetaK}
\theta_{\vec{k}} \, \equiv \, (2\pi)^{-3/2}\,\int \theta(\vec{x})\,e^{-i\vec{k}\cdot\vec{x}}\,d^3\vec{x}\,\,,
\eeq
and similary we can define the power spectrum of the velocity divergence
\beq
\label{eqn:Ptt(k)}
P_{\theta\theta}(k) \, = \, \left< |\theta_{\vec{k}}|^2\right> \, .
\eeq
In linear perturbation theory \GZ{the} ratio of the density power spectrum to the power spectrum of the velocity divergence scaled by the square of the growth rate should be unity. However on nonlinear and weakly nonlinear scales this ratio deviates from unity as in the nonlinear regime velocities grow more slowly than the linear perturbation theory prediction \citep[see, e.g.,][]{ps_ratio1,ps_ratio2}. Thus we would expect that the potential effects induced by the $f(R)$ fifth force in density and
divergence power spectra may differ from one another in the nonlinear and weakly nonlinear regimes.

\subsection{The structure of the density and velocity fields}

\label{subsect:patterns}

We start our analysis by comparing the visual impression of the density and velocity divergence fields obtained for our GR and $f(R)$ simulations. To measure the fields sampled from the distribution of dark matter (DM) particle positions and velocities we use the DTFE code of \citet{cv2011} \citep[see also][]{sv2000,vs2009}.
The Delaunay tessellation has the advantage that the velocity divergence field computed in this way is volume averaged,
as required by calculation, rather than mass averaged. 
%it also properly takes care of the empty cells where no particles
%can be found and thus zero velocity field would have been assumed in direct assignment. 
The DTFE code also avoids the problem of empty cells, where no particles can be found, which can arise in direct assignment methods to measure the velocity field \citep[see, e.g.,][]{ps2009}. Hence the velocity divergence power
spectrum measured in this way is unbiased and has better noise properties than is the case for standard interpolation methods that deal
with mass weighted velocities.

In Fig.~\ref{fig:dens_compare} we plot thin slices ($0.5\hmpc$) from the $z=0$ DM density field
in the $250\hmpc$ box. The panels show the GR model (top-left), the F6 (top-right) and the F5, F4 models (bottom row from left to right).
We note the large-scale structures and the patterns of the cosmic web are the same in all panels.
This is not unexpected as these simulations started from the same initial conditions, thus they share the same phases.
However, the delicate effects of the $f(R)$ fifth force can be observed on small scales, where some of the density field features
like filaments and clusters appear thicker in $f(R)$ universes. This is accompanied by deeper density dips in voids.

The picture becomes even more interesting when we look at the velocity divergence fields, which are plotted in Fig.~\ref{fig:divv_compare}. The slices shown correspond to the same regions plotted in Fig.~\ref{fig:dens_compare} from the $B=250h^{-1}$Mpc box. Here we plot only the GR and F4 models (for which the effect of the fifth force is
strongest) for three distinct epochs of cosmic evolution: $a=0.5$ (the bottom row), $a=0.7$ (the middle row) and $a=1.0$
(the top row). From the plot we observe that the differences between the GR and F4 models are very small at earlier stages
of evolution (the $a=0.5$ case), but they become very prominent as we move towards the present day ($a=1$). We 
note in particular that the velocity fields around and inside filaments and clusters are characterised by higher divergence. It is clear from this figure that, in the cosmic webs (clusters and filaments) of both the GR and and $f(R)$ cosmology, the velocity divergence shows larger deviations between the two cosmologies compared to the density field.
%\textit{It est} this feature of the $f(R)$ universe cosmic web (clusters and filaments) when compared to the fiducial model,
%appear much thicker in the velocity divergence field than what we could see in the density fields. \GZ{I don't understand this statement. Velocity field and density field are not comparable, even visually.}

\GZ{Note that the} velocity divergence field is positive in voids because matter flows from the void centres and the flow increases near the edge closer to higher density regions. Because the fifth force speeds up the matter flow, the velocity divergence in voids is larger in F4 than \GZ{that} in GR, and this is clearer at earlier times (note the difference in the colours of the two bottom panels). The trend is reversed around the clusters because matter flows inwards here, and again the magnitude of $\theta$ is larger for F4 due to the fifth force, which explains why the filaments appear to be thicker for F4 in Fig.~\ref{fig:divv_compare}. Inside clusters and filaments, the velocity divergence becomes positive again, as noted by \citet{ps2009}. \GZ{This is because of the virialisation: after the halo is formed, the particles stop falling into the potential wells but circle around the halo center, which can in principle make the velocity divergence positive.}

In the following sections we will precisely quantify and analyse these effects by studying both the matter and the velocity divergence power spectra in all $\Lambda$CDM and $f(R)$ cosmologies.

\begin{figure*}
 \includegraphics[width=160mm]{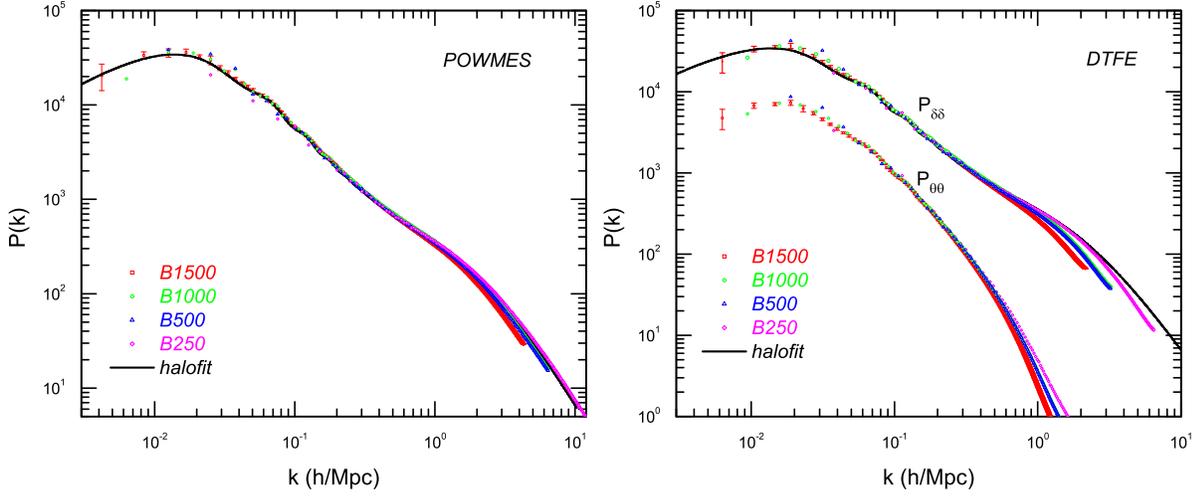}
\caption{(Colour Online) The power spectra of the $\Lambda$CDM model from different simulation boxes (symbols as explained in the legend; `Bxxx' means that the box size is xxx$h^{-1}$Mpc) compared to the {\tt HALOFIT} (black solid curve). {\it Left panel}: the matter power spectra measured using {\tt POWMES}. {\it Right panel}: the matter (upper symbols) and velocity divergence (lower symbols) power spectra measured from the DTFE-constructed density and velocity divergence fields.
%\WH{I strogly suggest to plot here power per-octave $k^3/2\pi^2 P(k)$ instead of just $P(k)$ as the former strobgly help to appriciate the diffriencies at small scales}
}
\label{fig:GR_z0_Pk}
\end{figure*}

\subsection{Measurement of power spectra}

\label{subsect:pk_meas}

The matter power spectrum has been measured from all the simulations listed in Table~\ref{table:simulations} using two codes: {\tt POWMES} \citep{powmes} and our own code that uses fields obtained from the DTFE method, which rely on different algorithms. {\tt POWMES} constructs the density field on a regular grid by direct particle assignment, while DTFE first samples the density and velocity divergence fields using Delaunay tessellation and then interpolates onto a regular grid. {\tt POWMES} attempts to correct for the impact of the scheme used to assign particles to the FFT grid, whilst we do not attempt any such correction with the code using the DTFE method.

The grid that is used for the power spectra measurement is chosen to have the same resolution as the domain grid in the simulations. For example, for the $L_{\rm box}=1.5h^{-1}$Gpc and $1.0h^{-1}$Gpc simulations the FFT grid has $1024^3$ cells. Only the results of the $L_{\rm box}=1.5h^{-1}$Gpc simulations have error bars, which show the scatter amongst all six realisations.

To show the accuracy of our simulations and the power spectrum codes, we plot the measured matter power spectra $P_{\delta\delta}$ from the $\Lambda$CDM simulations against the {\tt HALOFIT} \citep{halofit} prediction in Fig.~\ref{fig:GR_z0_Pk}. The {\tt HALOFIT} result is obtained using the publicly available {\tt CAMB} code \citep{camb}, assuming the same cosmological parameters as used in the simulations, and is used here simply as a reference.

The left panel of Fig.~\ref{fig:GR_z0_Pk} plots the $N$-body results measured using {\tt POWMES}, and the {\tt HALOFIT} power spectrum is plotted as a black solid line. It can be seen that the two agree very well over a wide range of length scales. For example, the matter power spectrum from the $L_{\rm box}=1.5h^{-1}$Gpc simulation is accurate for $0.004<k/(h/{\rm Mpc})<1$. This is of course as expected given that the {\tt RAMSES} code has been tested in many ways, and it gives us some reassurance about the {\tt ECOSMOG} gravity solver. In $f(R)$ gravity simulations, however, this relationship should only be used as a rough guide with caution, as we shall explain below in Sect.~\ref{subsect:pk_res}.

The right panel of Fig.~\ref{fig:GR_z0_Pk} shows the corresponding spectra measured from the DTFE-constructed density and velocity fields. The results also agree with {\tt HALOFIT} very well, especially on large scales. Note that in this case $P_{\delta\delta}$ starts to deviate from {\tt HALOFIT} at smaller $k$. This is due primarily to the lack of any correction in this estimate
for the effects of the scheme used to interpolate the smoothed density field onto the FFT grid \citep[see][]{j2005,powmes}. 
%Fig. 3 shows that the DTFE estimate of the power spectrum is
%accurate to $\approx 0.3 k_{\rm Nyq}$ and is damped beyond this scale.
%
%whilst there are certain tricks to push the good agreement to even larger $k$ \citep[for more details see][]{powmes}.

\begin{figure}
\includegraphics[width=80mm]{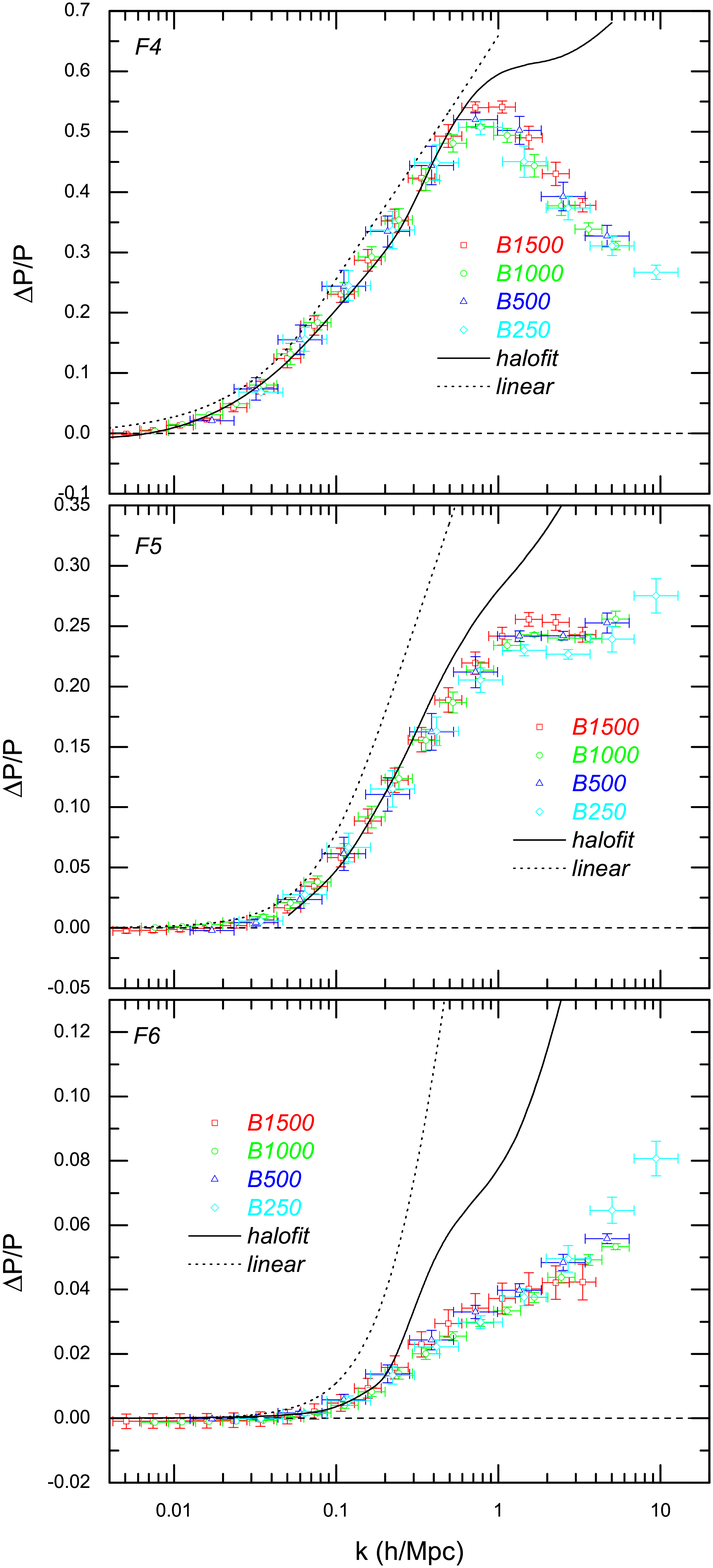}
\caption{(Colour Online) The relative difference between the matter power spectra of the $f(R)$ and $\Lambda$CDM simulations at $z=0$. The results have been binned along the $k$-axis as described in the text. `Bxxxx' in the legend means that the simulation box size is xxxx$h^{-1}$Mpc, and the horizontal dashed line is identically zero. The top to bottom panels show respectively results for models F4, F5 and F6. The black dotted and solid curves are respectively the predictions using linear perturbation theory and {\tt HALOFIT} \citep{halofit}.}
\label{fig:PK_diff_z0}
\end{figure}

\subsection{Resolution issues in $f(R)$ simulations}

\label{subsect:pk_res}

%The shape of matter power spectrum could be a unique character of the underlying gravitational theory or properties of the dark energy.
The shape of the matter power spectrum is sensitive to changes in the cosmological model assumed and as such it may be a sensitive probe of the underlying theory of gravity or the properties of dark energy. 
Here, we are mostly interested in the shape of $\Delta P_{\delta\delta}/P_{\delta\delta}$, where $\Delta P_{\delta\delta}$ is the difference between the matter power spectra for $f(R)$ gravity and $\Lambda$CDM, defined as
\beq
\label{eqn:pw_diff}
{\Delta P_{\delta\delta}\over P_{\delta\delta}} \equiv {P_{\delta\delta}^{f(R)}(k)\over P_{\delta\delta}^{LCDM}(k)} - 1\,.
\eeq
To make the plots clearer, we have rebinned. %separated the results into a number of bins in the $k$-axis. 
The data points are at the centres of the bins, and the average value of $P(k)$ and error bars are computed as follows: for the $L_{\rm box}=1.5h^{-1}$Gpc simulations, the $P(k)$ value is the average over all points in a given bin in all six realisations and the error bar shows the scatter amongst all these points; for all other simulations, the same thing is done but only for points within a given bin in a single realisation.

The results for models F4/F5/F6 at $a=1$ are shown in Fig.~\ref{fig:PK_diff_z0}, and different symbols are used to denote different simulation box sizes. We can see that the different symbols overlap with each other quite well, especially on large scales. On small scales, the large-box simulations predict slightly larger difference in $P_{\delta\delta}$, but the difference is, in general, quite small, at least at this particular cosmic time. In Fig.~\ref{fig:PK_diff_z0} we have also overplotted the results from linear perturbation calculation and {\tt HALOFIT}. The {\tt HALOFIT} results are obtained from the linear power spectra in $f(R)$ gravity in the fitting formulae obtained by \citep{halofit}. We have checked that on large scales the {\tt HALOFIT} result agrees with third-order perturbation theory \citep{kth2009} quite well, and both show better agreement with simulations compared with the linear perturbation theory. Indeed, linear theory breaks down on almost all scales where the $f(R)$ model deviates from GR, 
especially in the F4/F5 cases where the nonlinearity is stronger. On the other hand, the {\tt HALOFIT} gives a worse fit to simulation results in F6. This is because {\tt HALOFIT} is calibrated using GR simulations and it does not capture the effect of the chameleon mechanism in F6, which suppresses the deviation from $\Lambda$CDM on small scales.

The shape of $\Delta P_{\delta\delta}/P_{\delta\delta}$ for the three $f(R)$ models look quite different from each other. In the F6 case,
this increases all the way down to the smallest scales probed by the simulations; in F5, a small bump appears in between $k=1h/$Mpc and $k=2/$Mpc,
while after that $\Delta P_{\delta\delta}/P_{\delta\delta}$ increases on small scales; for F4, a single peak appears at $k\sim1h/$Mpc, and on
smaller scales $\Delta P_{\delta\delta}/P_{\delta\delta}$ simply decreases. In addition, the amplitude of $\Delta P_{\delta\delta}/P_{\delta\delta}$
increases with $|f_{R0}|$. These features were also seen by \citet{lz2009,lz2010} in their chameleon simulations, then \citet{zlk2011} in their small $f(R)$ simulations, and are confirmed here by our larger simulations; Recently, similar features have also been found in simulations of generalised dilaton and symmetron models \citep{bdlwz2012}\footnote{Other authors in their studies of cosmologies employing simpler forms of fifth force
have found similar features in the matter power spectra as we have observed here \citep[e.g.][]{rebel,rebel2}.}. This can lead to the conclusion that
in general the dynamics with a fifth force employing a specific spatial screening mechanism leaves a characteristic mark visible in statistics
of spatial clustering. A proper understanding of the features which appear in the power spectra of our models
can only be achieved when we have a clear picture of the time evolution of $\Delta P_{\delta\delta}/P_{\delta\delta}$, and for this
we plot the results of F4 and F5 at $a=0.3, 0.5$ and $0.7$ in Fig.~\ref{fig:Pdd_evol}.

Before going through the details of the time evolution of $\Delta P_{\delta\delta}/P_{\delta\delta}$, let us note that our previous observation, that the large-box simulations tend to overestimate $\Delta P_{\delta\delta}/P_{\delta\delta}$ on small scales, becomes more prominent at earlier times. As an example, at $a=0.5$ we find that $\Delta P_{\delta\delta}/P_{\delta\delta}$ has a peak value of $\sim70\%$ according to the $L_{\rm box}=1.5h^{-1}$Gpc simulation, while this value decreases for high-resolution simulations and drops to $\sim40\%$ for the $250h^{-1}$Mpc boxes.

One probable reason for this is the different force resolutions in these simulations. As the particles cluster to form structures, local over-densities grow and the chameleon effect starts to suppress the fifth force. If the force resolution is too low, the density field tends to be underestimated\footnote{The density in each cell is computed using a triangular-shaped cloud assignment scheme, and using larger cells means that the mass of a particle will be more widely spread and thus the density peaks lower.} and the fifth force overestimated, resulting in more clustering of matter in these simulations compared to the ones which have higher resolution.

Note that this resolution effect is a separate issue which happens for the calculation of the chameleon fifth force: with the same background density, putting a particle at the centre of a sphere with radius $R$ or spreading its mass uniformly in the sphere produce the same gravity at $R$, but the fifth forces at $R$ would be quite different in these two configurations. This has nothing to do with the resolution that is required by the usual gravity (Poisson equation) solver.

The implications of this result are:
\begin{enumerate}
\item Comparing the $\Lambda$CDM power spectrum with {\tt HALOFIT} as we did in Fig.~\ref{fig:GR_z0_Pk} does not automatically provides a useful guide as to the scales down to which the $f(R)$ simulations are reliable. The resolution effect on the fifth force solver is complicated; it depends on the model (\eg, compare F4 and F5 at $a=0.5$) as well as on the redshift (\eg, compare F4 at $a=0.5$ and F4 at $a=1.0$).
\item One has to be careful in choosing the right resolution to obtain accurate results in the $f(R)$ simulations. The most straightforward way would be to run ever higher-resolution simulations and see where the results start to disagree with each other.
\end{enumerate}

As an illustration, for F4 the $L_{\rm box}=1.5h^{-1}$Gpc simulations can be trusted down to $k\sim0.4h/$Mpc at $a=0.5$, $k\sim1h/$Mpc at $a=1.0$, while its qualitative predictions can be trusted down to even smaller scales. Evidently, such a box is not good enough to study small structure in the simulations, but is sufficient to study large-scale properties, such as redshift-space distortions \citep{jblkz2012}.

\begin{figure*}
\includegraphics[width=170mm]{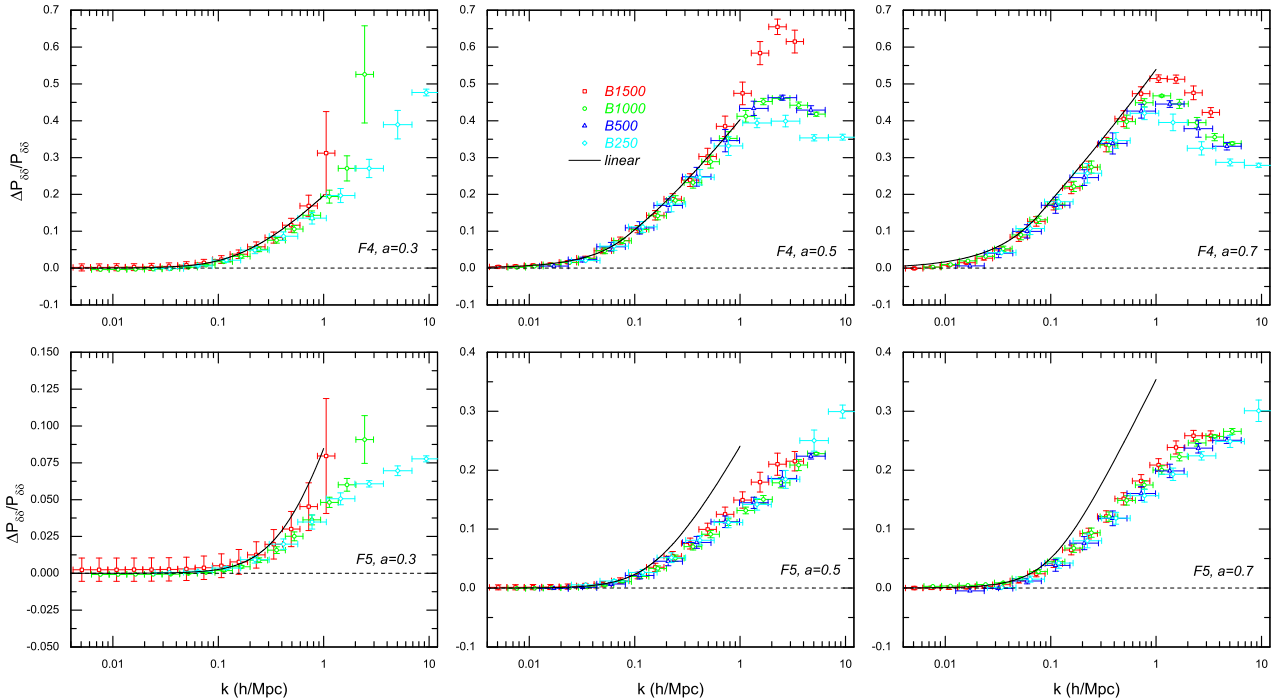}
\caption{(Colour Online) The time evolution of $\Delta P_{\delta\delta}/P_{\delta\delta}$ for models F4 (upper panels) and F5 (lower panels). The left, middle and right panels are respectively the results at $a=0.3$, $0.5$ and $0.7$ (the $a=1$ results are shown in Fig.~\ref{fig:PK_diff_z0}). `Bxxxx' in the legend means that the simulation box size is xxxx$h^{-1}$Mpc, the horizontal dashed line is identically zero and the solid black curve is the linear perturbation prediction which agrees with the simulations better at earlier times.}
\label{fig:Pdd_evol}
\end{figure*}

\begin{figure}
 \includegraphics[angle=-90,width=80mm]{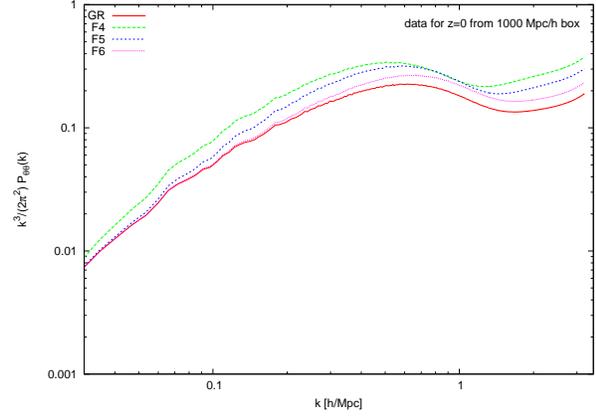}
\caption{(Colour Online) The patterns of the small-scale tails of the velocity divergence power spectrum per octave $k^3/2\pi^2 P_{\theta\theta}(k)$ for the different cosmologies measured at $z=0$ from the $B=1000h^{-1}$Mpc simulation box.
}
\label{fig:ptt_zoom}
\end{figure}

\begin{figure*}
\includegraphics[width=170mm]{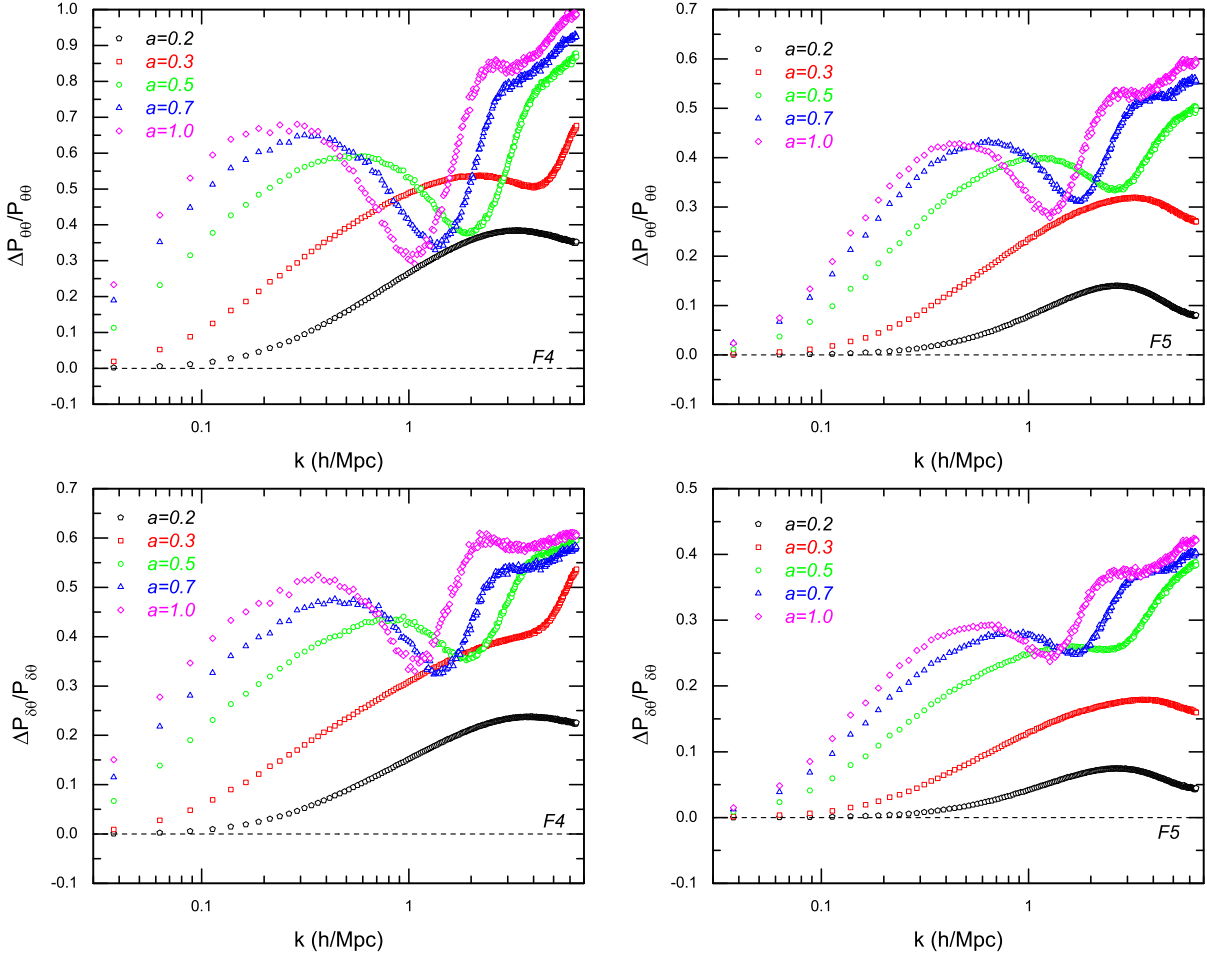}
\caption{(Colour Online) The time evolution of $\Delta P_{\theta\theta}/P_{\theta\theta}$ in F4 (upper left panel) and F5 (upper right model), along with the time evolution of $\Delta P_{\delta\theta}/P_{\delta\theta}$ for model F4 (lower left panel) and model F5 (lower right panel). In all panels at $k=0.1h/$Mpc the symbols are for times $a=0.2, 0.3, 0.5, 0.7$ and $1.0$ from bottom to top. All results are measured from the DTFE-constructed density and velocity divergence fields for the $L_{\rm box}=250h^{-1}$Mpc simulations.}
\label{fig:Pdt_evol}
\end{figure*}

\subsection{Time evolution of the spectra}

\label{subsect:pk_evol}

Let us now discuss on the time evolution of $\Delta P_{\delta\delta}/P_{\delta\delta}$. From Figs.~\ref{fig:PK_diff_z0} and \ref{fig:Pdd_evol} we can see that as the Universe evolves, not only the magnitude but also the shape of $\Delta P_{\delta\delta}/P_{\delta\delta}$ changes. Taking the F4 model as an example: at $a=0.3$ $\Delta P_{\delta\delta}/P_{\delta\delta}$ increases as one goes to smaller scales (at least until $k\sim10h$/Mpc); at $a=0.5$ a peak develops at $k_{\rm peak}\sim2h$/Mpc while $\Delta P_{\delta\delta}/P_{\delta\delta}$ decreases for $k>k_{\rm peak}$; then the peak of $\Delta P_{\delta\delta}/P_{\delta\delta}$ shifts towards larger scales with $k_{\rm peak}\sim1h$/Mpc at $a=0.7$ and $k_{\rm peak}\sim0.9h$/Mpc at $a=1$. The F5 model behaves similarly but the peak only develops at $a\sim1$. Likewise, the F6 model does not develop any peak in $\Delta P_{\delta\delta}/P_{\delta\delta}$ by $a=1$.

For the velocity divergence power spectrum $P_{\theta\theta}$ and the cross power spectrum $P_{\delta\theta}$, we use those measured from the  $L_{\rm box}=250h^{-1}$Mpc simulations. The result is not sensitive to the mass resolution, though this box size is a bit too small, which means that the measured $P_{\theta\theta}$ can be $\sim5-10\%$ higher \citep{ps2009}. However, we are interested in the qualitative behaviour rather than accurate measurement of $P_{\theta\theta}$, and this box enables us to go to smaller scales.

%To further appriciate the patterns in $P_{\theta\theta}(k)$ that developes at small scales during nonlinear evolution
%we also plot a zoom-in into $0.03\leq k/(Mpc/h) \leq 3$ region. To make the diffriencies at small-scales better visible we plot there the power per octave $k^3/(2\pi^2) P_{\theta\theta}(k)$ rather then pure $P_{\theta\theta}$. We can see that the peak-dip-peak pattern appearing for $k>0.1h/Mpc$ is a general feature of the nonlinear velocity-divergence
%power spectrum as it is present in all our models (including the fiducial $\Lambda$CDM). It appears so tha the features seen in the Fig. \ref{fig:Pdt_evol} are the reflecion
%of the general peak-dip-peak pattern strongly enhanced by the fifth force of the $f(R)$ gravity.
\BLED{Fig.~\ref{fig:ptt_zoom} shows the behaviour of $P_{\theta\theta}$ on small scales ($0.03\leq k/(h/{\rm Mpc}) \leq 3$) in the different cosmologies. To make the curves clearer we have plotted $k^3/(2\pi^2) P_{\theta\theta}(k)$ instead of $P_{\theta\theta}$. From this plot we can see that 
\begin{enumerate}
\item there is a peak-dip-peak pattern on small scales, which agrees with what we have seen in the velocity divergence field in Fig.~\ref{fig:divv_compare},
\item not only the power spectrum is enhanced by the fifth force, but the peaks and dip also shift towards larger scales as the fifth force becomes stronger. This implies that the peak-dip-peak pattern observed in the plots develops as structures form, and could possibly be related to the characteristic scales of the structure formation at a given time, as we will show in the next subsection.
\end{enumerate}}

Fig.~\ref{fig:Pdt_evol} illustrates the time evolution of $\Delta P_{\theta\theta}/P_{\theta\theta}$ (upper panels) and $\Delta P_{\delta\theta}/P_{\delta\theta}$ (lower panels) for models F4 (left column) and F5 (right column). We can see some interesting features in these plots. Taking $\Delta P_{\theta\theta}/P_{\theta\theta}$ of the F4 model as an example, at early times (e.g., $a=0.2$) this ratio increases with $k$ until small scales; a dip then develops e.g., at $k_{\rm dip}\sim4h$/Mpc at $a=0.3$ and the dip shifts towards larger scales at late times. Meanwhile, a peak appears at $k_{s}<k_{\rm dip}$, while for $k>k_{\rm dip}$ $\Delta P_{\theta\theta}/P_{\theta\theta}$ goes up again. Furthermore, a second and minor dip could develop immediately to the right of $k_{\rm dip}$. The features and evolution pattern for $\Delta P_{\delta\theta}/P_{\delta\theta}$ are very similar to $\Delta P_{\theta\theta}/P_{\theta\theta}$, both being significantly larger than $\Delta P_{\delta\delta}/P_{\delta\delta}$ \citep{jblkz2012}: this implies that local measurements of the velocity field \citep{pbp2012,p_vel1,p_vel2,p_vel3,p_vel4} can indeed be a good probe of modified gravity. 

Of course, the complicated shape and evolution of $\Delta P_{\theta\theta}/P_{\theta\theta}$ must come from the chameleon fifth force, and we shall present an explanation of this below.

The above evolution pattern suggests that the time evolution of $\Delta P_{\delta\delta}/P_{\delta\delta}$ and  $\Delta P_{\theta\theta}/P_{\theta\theta}$ for F5 (F6) is just a postponed version of that for F4. If this is the case, then there is a natural explanation for this, namely that the whole evolution pattern is an effect of the fifth force, which is suppressed until later times for smaller values of $|f_{R0}|$. We will describe this in more detail in the next subsection. Note that this time-shifting effect can also be seen in the linear perturbation results for $\Delta P_{\delta\delta}/P_{\delta\delta}$ and $\Delta P_{\theta\theta}/P_{\theta\theta}$ shown in Fig.~\ref{fig:linear}.

\begin{figure*}
	\subfigure{%
		\includegraphics[clip, width=90mm]{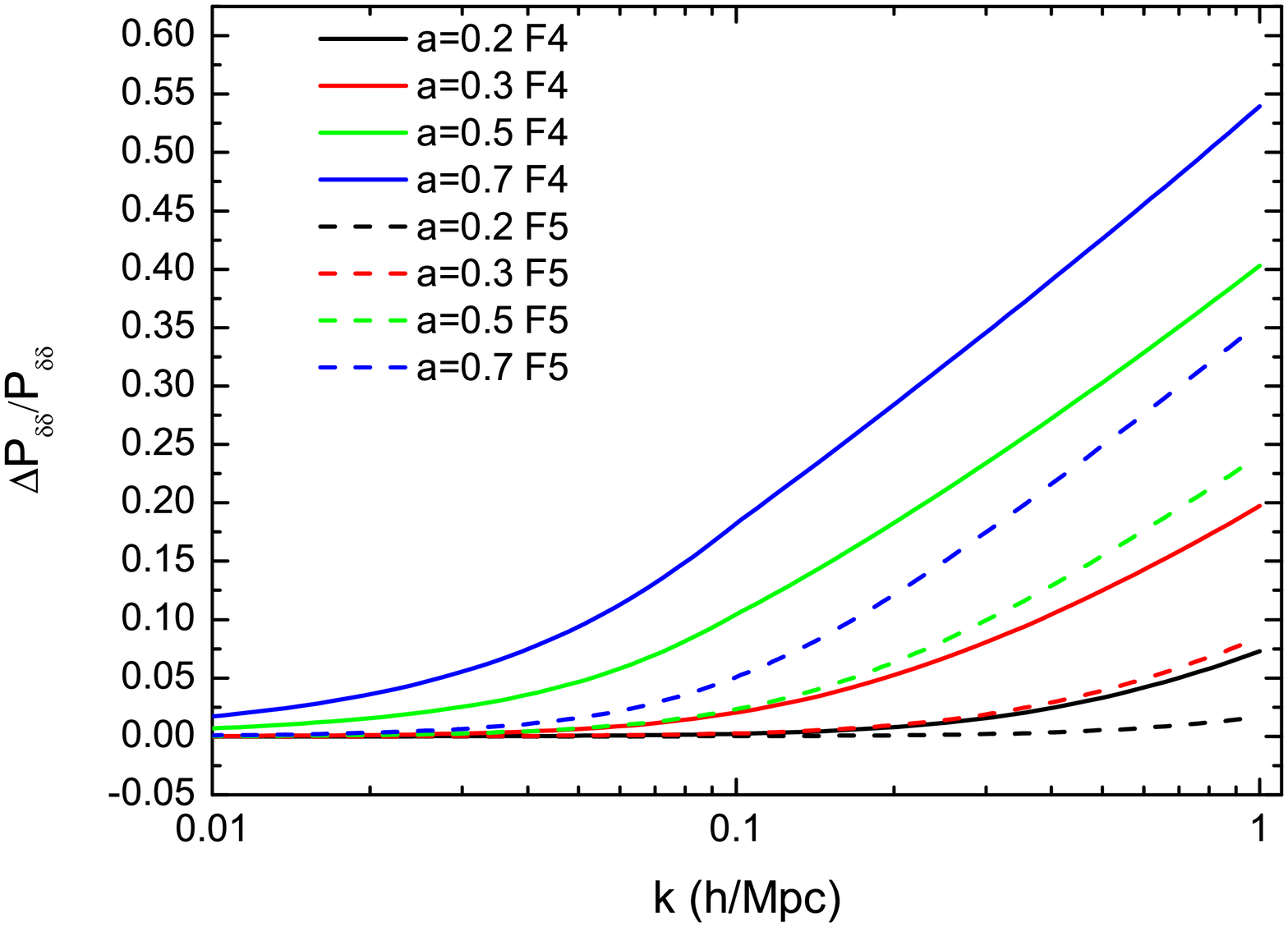}}%
	\subfigure{%
		\includegraphics[clip, width=90mm]{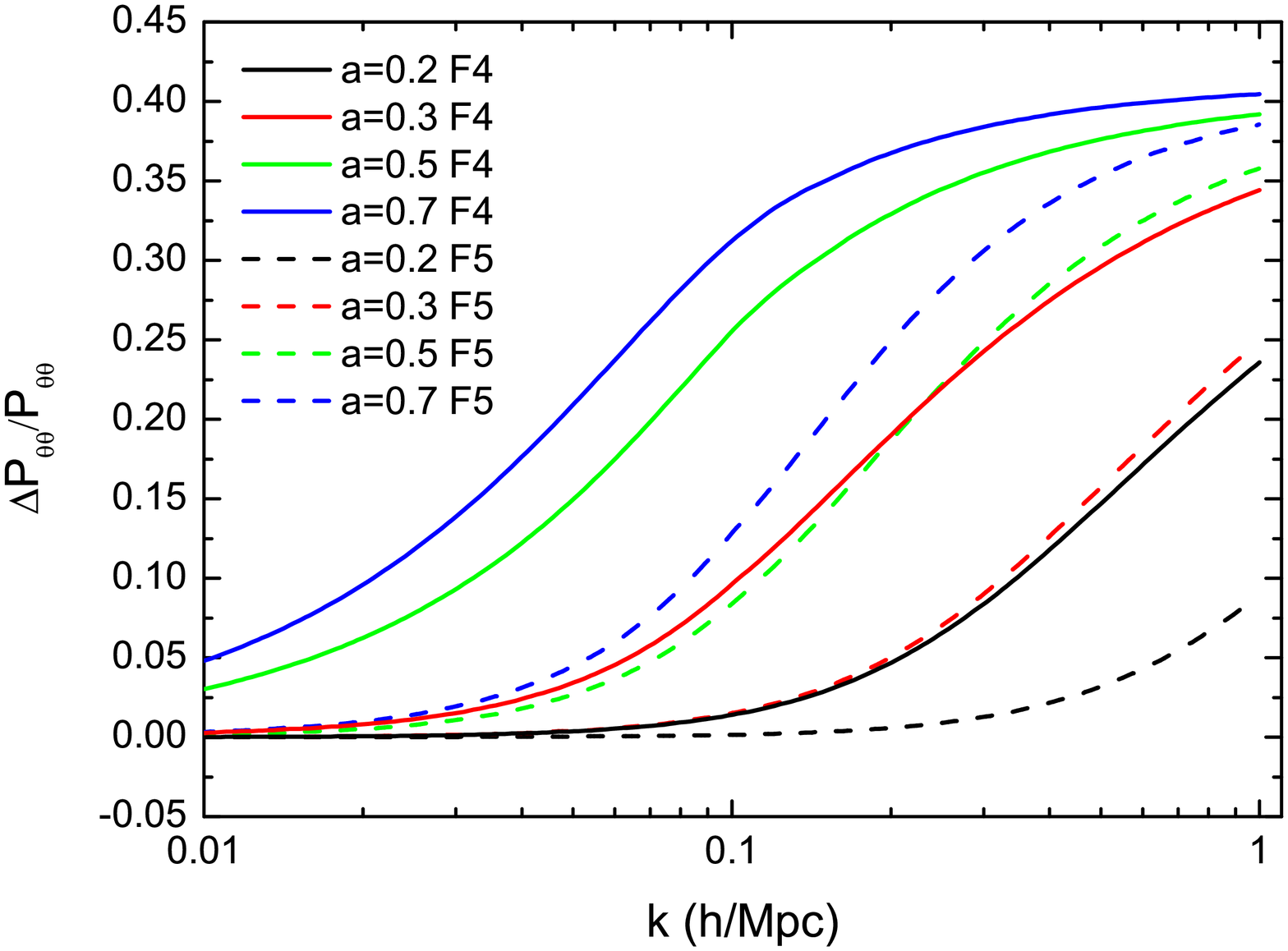}}%
	\caption{(Colour Online) Linear theory predictions for the time evolution of $\Delta P_{\delta\delta}/P_{\delta\delta}$ (left panel) and $\Delta P_{\theta\theta}/P_{\theta\theta}$ (right panel) for F4 (solid) and F5 (dashed) at different expansion factors, as indicated by the labels.}
	\label{fig:linear}
\end{figure*}

\subsection{A tale of two universes}

\label{subsect:pk_theory}

Suppose that there are two universes which are completely identical except for the underlying gravity. In universe I, standard general relativity applies and in universe II, the $f(R)$ gravity applies. The evolution of structure is the same in the two universes up to quite late times, say $z=49$ which is the starting time of our experiments, since the fifth force in $f(R)$ gravity is vastly suppressed until then.

The subsequent evolution can be divided into several stages:

1) Stage (a), the fifth force in $f(R)$ gravity begins to affect increasingly larger scales, starting from the smallest one. This speeds up the flow of matter, making the smallest structures form earlier through collapse in universe II than in universe I (such a boost in the rate of the formation of hierarchical structures was reported for a general fifth-force models by \citet{lz2009,lz2010,rebel3} and \citet{lb2011}). As a result, one can see that both $\Delta P_{\delta\delta}/P_{\delta\delta}$ and $\Delta P_{\theta\theta}/P_{\theta\theta}$ increase towards small scales.

2) Stage (b), the small structures have formed, and this corresponds to shell crossing in the spherical collapse model in universe II, while in universe I the same structure is still forming. The velocity divergence inside the collapsed regions becomes less negative and then positive, during which process its magnitude is smaller than in the collapsing regions [cf.~Fig.~\ref{fig:divv_compare}]. Although this is the effect for an individual structure, we would expect to see this statistically (and same for the discussions below), i.e., in the power spectrum, because structures form earlier in F4 in general, and as a result a dip starts to appear in $\Delta P_{\theta\theta}/P_{\theta\theta}$ on the scale roughly corresponding to the collapsed structure. During this stage, $\Delta P_{\delta\delta}/P_{\delta\delta}$ still follows the pattern of stage (a).

3) Stage (c), the same small structure collapses and forms in universe I as well. Inside the structure, the deepening of the total gravitational potential (with the fifth force contributing) in universe II makes matter move faster than it does in universe I. This fact is reflected  as a continued increase in $\Delta P_{\theta\theta}/P_{\theta\theta}$ on scales larger than where the dip first appears. Meanwhile, the dip develops into a valley and moves towards larger scales, because structures form hierarchically and larger ones form later. During this stage, $\Delta P_{\theta\theta}/P_{\theta\theta}$ also continues growing on scales much larger than the collapsing regions. A small peak could appear on scales immediately smaller than that corresponds to the valley, because the divergence field crosses zero, making $P_{\theta\theta}$ smaller for GR; this is probably why there is a second and smaller dip in $\Delta P_{\theta\theta}/P_{\theta\theta}$.

The evolution of $\Delta P_{\delta\delta}/P_{\delta\delta}$ in stage (c) is more complicated. During the first part, substage (ci), the small scale clustering is continuously boosted by the fifth force in universe II, which means that $\Delta P_{\delta\delta}/P_{\delta\delta}$ keeps growing towards small scales. Meanwhile, a bump starts to appear on scales roughly corresponding to the collapsing regions, reflecting the enhanced and earlier formation of larger structures in universe II, probably as well as the fact that the growth of $\Delta P_{\delta\delta}/P_{\delta\delta}$ on scales smaller than that corresponding to the bump is slowed down by the increased velocity dispersion in the structures in universe II \citep{lb2011}.

This continues into substage (cii), during which the bump in $\Delta P_{\delta\delta}/P_{\delta\delta}$ develops into a peak and shifts towards larger scales (at the same pace as the valley in $\Delta P_{\theta\theta}/P_{\theta\theta}$ shifts leftwards). At the same time $\Delta P_{\delta\delta}/P_{\delta\delta}$ goes down on small scales, because of the higher velocity dispersion inside halos in universe II.

The above evolution history of the shapes of $\Delta P_{\delta\delta}/P_{\delta\delta}$ and $\Delta P_{\theta\theta}/P_{\theta\theta}$ depends on the properties of the fifth force (\eg, it grows in time), and could in principle be a unique feature of the chameleon-type modified gravity theories. According to this picture, the different $f(R)$ models studied in this paper should follow the same evolutionary path, but as the fifth force becomes non-negligible in different eras depending on the value of $|f_{R0}|$, at any given time the evolution is at different stages for the different models.

As an illustration, at $a=1.0$ F4 has reached stage (cii), F5 reached stage (ci) and F6 somewhere between stages (b) and (ci). At $a=0.7$, F4 and F5 are in stages (cii) and (b) respectively. At $a=0.5$, F4 has just left stage (ci) while F5 is still in stage (b).

Note that the decrease of $\Delta P_{\delta\delta}/P_{\delta\delta}$ towards small scales in the F4 model does {\it not} reflect the fact that the fifth force is suppressed in small systems such as clusters and galaxies, although the latter is true. 

Based on these observations, we can give a rough estimate of the scales where the peak in $\Delta P_{\delta\delta}/P_{\delta\delta}$ and the dip in $\Delta P_{\theta \theta}/P_{\theta\theta}$ appear. We define the variance of linear density fluctuations smoothed by a Gaussian filter as
\begin{equation}
\sigma^2(R,z) = \int \frac{k^3 P_L(k)}{2 \pi^2} \exp(-k^2 R^2) d \ln k,
\end{equation}
where $P_L(k)$ is the linear power spectrum. The characteristic mass of halos $M_*$ is defined by matching the variance of the linear density fluctuation to the threshold density for collapse, $\delta_c$,
\begin{equation}
\sigma(R_*,z) =\delta_c,
\end{equation}
where $M_* = 4 \pi R_*^3 \bar{\rho}_m /3$. We expect that on small scales at $k > k_* \equiv
R_*^{-1}$, the power spectrum is significantly affected by collapsed objects. Using the critical density obtained by the spherical collapse model, $\delta_c=1.673$ for the $\Lambda$CDM model and $\delta_c=1.692$ \citep{sloh2009} for F4\footnote{{Note that here we have assumed that $\delta_c$ for F4 is scale-independent as in GR: this is obtained by rescaling the Newton constant by $4/3$ everywhere and the chameleon effect is neglected.}}, the characteristic scales are obtained as $k_*=1.05 h$Mpc$^{-1}$ in $\Lambda$CDM and $k_*=0.72 h$Mpc$^{-1}$ in F4 at $a=1$. At $a=0.5$, these scales are give by $k_*=3.39 h$Mpc$^{-1}$ in $\Lambda$CDM and $k_*=2.2h$Mpc$^{-1}$ in F4. The characteristic scale $k_*=R_*^{-1}$ is always smaller in $f(R)$ gravity as the nonlinearity is stronger than $\Lambda$CDM. From the above arguments, we expect that the peak in  $\Delta P_{\delta\delta}/P_{\delta\delta}$ and the dip in $\Delta P_{\theta \theta}/P_{\theta\theta}$ appear roughly at $k_*$ (F4) $< k < k_*$ ($\Lambda$CDM), as on these scales, collapsed objects are already formed in F4, but they are still collapsing in $\Lambda$CDM. From Figs.~\ref{fig:PK_diff_z0}, \ref{fig:Pdd_evol}, \ref{fig:Pdt_evol}, we can see that these scales are roughly consistent with the scales where the peak and the dip appear. We should emphasize that these scales only give qualitative estimates of scales where collapsed objects are important and it is necessary to study the formation of halos in detail to make more precise predictions.

\section{Summary and conclusions}

\label{sect:con}

To summarise, in this paper we have studied the shape and evolution of the matter and velocity divergence power spectra in the $f(R)$ gravity model, with the aid of a number of high-resolution $N$-body simulations. For this we have worked with the $f(R)$ Lagrangian proposed by \citet{hs2007}, fixing one of the two free parameters (namely setting $n=1$ in Eq.~\ref{eq:hs}). This leaves us with only one free parameter $|f_{R0}|$, which is the present-day value of $f_R$ in the cosmological background. The value of $|f_{R}|$ controls the strength of the chameleon mechanism: the smaller $|f_{R}|$ is, the stronger the chameleon effect becomes and the weaker the deviations from general relativity. Because $|f_R|$ increases with time overall, a larger value of $|f_{R0}|$ means that the fifth force becomes unscreened at an earlier time.

We have run a series of $N$-body simulations to study the formation of cosmic structures in selected $f(R)$ models using the {\tt ECOSMOG} code. To assess all possible resolution and finite box effects we make sure that our simulations cover a wide range of length and mass scales. On very large scales, the matter power spectrum of $f(R)$ gravity is found to be the same as that of the $\Lambda$CDM paradigm, since these scales are well beyond the range of the fifth force. On small scales, the matter power spectrum develops nontrivial shapes, depending on the value of $|f_{R0}|$ and time. We stress that linear perturbation theory is a bad approximation even on large scales, especially for the cases with $|f_{R0}|=10^{-5}$ and $10^{-6}$, in which the chameleon effect is strong and the scalaron equation is highly nonlinear. This implies that one should be cautious about forecasts made for modified gravity theories based on linear perturbation theory calculations. In general full nonlinear numerical simulations are needed.

The most challenging part of the $f(R)$ simulation (and modified gravity simulation in general) is that the fifth force becomes weak in high density regions, where higher resolution is needed. We have seen in \S~\ref{subsect:pk_res} that if the mass and force resolution is not high enough, the amplitude of density peaks could be underestimated and the magnitude of the fifth force overestimated, causing significant errors in the simulations.

The peculiar velocity field in the $f(R)$ gravity is more affected by the presence of the fifth force than the density field. Indeed, the velocity divergence power spectrum of the $f(R)$ gravity can differ from that of $\Lambda$CDM by twice as much as the difference in the matter power spectrum ($\sim100\%$ versus $\sim50\%$ for F4 and $\sim60\%$ versus $\sim30\%$ for F5). Furthermore, the shape and evolution pattern of the velocity divergence power spectrum, although also dependent on $|f_{R0}|$ and time, can be very different from those of the matter power spectrum. The large effect of modified gravity on the velocity divergence power spectrum, especially on small scales, implies that the motion of particles and thus the dynamical state of the halos can be very different in modified gravity theories. Galaxy rotational curves, for example, can be modified and this effect is important when interpreting observational data. The spin of dark matter halos, especially in low-density regions, can also be significantly faster \citep{lzlk2012}.

The dependencies of the matter and velocity divergence power spectra on $|f_{R0}|$ and time can be simplified if one understands them as the dependency on a single quantity -- the fifth force. We have shown that the shapes of the power spectra for different $|f_{R0}|$ actually evolve on the same path, but for models with smaller $|f_{R0}|$ the fifth force is suppressed until later times and the whole evolution is delayed. For example, the F5 $\Delta P_{\theta\theta}/P_{\theta\theta}$ at $a=1$ looks like the F4 results at $a=0.7$. 

We have presented an explanation of the shape and evolution of the power spectra based on this observation, according to which the valley and peaks in $\Delta P_{\theta\theta}/P_{\theta\theta}$ appear as a result of the fact that structures form earlier in $f(R)$ gravity than they do in the $\Lambda$CDM model. This also explains the observation in \citet{zlk2011} (and also \citet{lz2009,lz2010} for other chameleon-type models) that $\Delta P_{\delta\delta}/P_{\delta\delta}$ first increases as $k$ increases and later develops a peak at the $k$ corresponding to the size of dark matter halos. 

In this paper we have only focused on the theoretical aspects of the power spectra in $f(R)$ gravity. The qualitative results here are expected to be quite general, and according to the theoretical picture similar things would be found in other modified gravity theories with screening mechanisms, such as the symmetron and dilaton models or the ReBEL model \citep{NGP}. Our analysis could be generalised to those models, and also connections could be made to observations by, for example, considering the weak lensing shear spectrum etc., to place constraints on the parameter $|f_{R0}|$. These issues will be left to future work.

\section*{Acknowledgments}

BL is supported by the Royal Astronomical Society and Durham University.
WAH acknowledges supports from Polish National Science Center (grant No.~DEC-2011/01/D/ST9/01960) and ERC Advanced Investigator grant (C.~S.~Frenk), COSMIWAY.
KK and GBZ acknowledge support from the STFC (grant No.~ST/H002774/1), and KK is also supported by an ERC Starting Grant and the Leverhulme Trust. 
EJ is supported by a grant from the Simons Foundation (award No.~184549), the Kavli Institute for Cosmological Physics at the University of Chicago (grants NSF PHY-0114422 and NSF PHY-0551142) and an endowment from the Kavli Foundation and its founder Fred Kavli. 
We thank Shaun Cole and Carlos Frenk for useful comments and discussions.
The simulations for this paper were performed on the ICC Cosmology Machine, which is part of the DiRAC Facility jointly funded by STFC, the Large Facilities Capital Fund of BIS, and Durham University.
We thank Lydia Heck for technical support.

\bibliographystyle{mn2e}
\bibliography{pkfr}

\begin{thebibliography}{}

\bibitem[\protect\citeauthoryear{{Adelberger}, {Heckel} \&
  {Nelson}}{{Adelberger} et~al.}{2003}]{gr_test4}
{Adelberger} E.~G.,  {Heckel} B.~R.,    {Nelson} A.~E.,  2003, Annual Review of
  Nuclear and Particle Science, 53, 77

\bibitem[\protect\citeauthoryear{{Amendola}}{{Amendola}}{2000}]{a2000}
{Amendola} L.,  2000, \prd, 62, 043511

\bibitem[\protect\citeauthoryear{Appleby \& Battye}{Appleby \&
  Battye}{2007}]{ab2007}
Appleby S.~A.,  Battye R.~A.,  2007, Phys.Lett., B654, 7

\bibitem[\protect\citeauthoryear{{Bertotti}, {Iess} \& {Tortora}}{{Bertotti}
  et~al.}{2003}]{gr_test1}
{Bertotti} B.,  {Iess} L.,    {Tortora} P.,  2003, \nat, 425, 374

\bibitem[\protect\citeauthoryear{{Biswas} \& {Notari}}{{Biswas} \&
  {Notari}}{2008}]{bn2008}
{Biswas} T.,  {Notari} A.,  2008, \jcap, 6, 21

\bibitem[\protect\citeauthoryear{{Brax}, {Davis}, {Li} \& {Winther}}{{Brax}
  et~al.}{2012}]{bdlw2012}
{Brax} P.,  {Davis} A.-C.,  {Li} B.,    {Winther} H.~A.,  2012, ArXiv:1203.4812
  [astro-ph.CO]

\bibitem[\protect\citeauthoryear{{Brax}, {Davis}, {Li}, {Winther} \&
  {Zhao}}{{Brax} et~al.}{2012}]{bdlwz2012}
{Brax} P.,  {Davis} A.-C.,  {Li} B.,  {Winther} H.~A.,    {Zhao} G.,  2012,
  arXiv preprint

\bibitem[\protect\citeauthoryear{{Brax}, {van de Bruck}, {Davis}, {Li} \&
  {Shaw}}{{Brax} et~al.}{2011}]{bbdls2011}
{Brax} P.,  {van de Bruck} C.,  {Davis} A.-C.,  {Li} B.,    {Shaw} D.~J.,
  2011, \prd, 83, 104026

\bibitem[\protect\citeauthoryear{{Brax}, {van de Bruck}, {Davis} \&
  {Shaw}}{{Brax} et~al.}{2008}]{bbds2008}
{Brax} P.,  {van de Bruck} C.,  {Davis} A.-C.,    {Shaw} D.~J.,  2008, \prd,
  78, 104021

\bibitem[\protect\citeauthoryear{{Brax}, {van de Bruck}, {Davis} \&
  {Shaw}}{{Brax} et~al.}{2010}]{bbds2010}
{Brax} P.,  {van de Bruck} C.,  {Davis} A.-C.,    {Shaw} D.~J.,  2010, \prd,
  82, 063519

\bibitem[\protect\citeauthoryear{Brookfield, van~de Bruck \& Hall}{Brookfield
  et~al.}{2006}]{bbh2006}
Brookfield A.~W.,  van~de Bruck C.,    Hall L.~M.,  2006, Phys.Rev., D74,
  064028

\bibitem[\protect\citeauthoryear{{Carroll}, {De~Felice}, {Duvvuri}, {Easson},
  {Trodden} \& {Turner}}{{Carroll} et~al.}{2005}]{cddett2005}
{Carroll} S.~M.,  {De~Felice} A.,  {Duvvuri} V.,  {Easson} D.~A.,  {Trodden}
  M.,    {Turner} M.~S.,  2005, \prd, 71, 063513

\bibitem[\protect\citeauthoryear{{Cautun} \& {van de Weygaert}}{{Cautun} \&
  {van de Weygaert}}{2011}]{cv2011}
{Cautun} M.~C.,  {van de Weygaert} R.,  2011, ArXiv:1105.0370 [astro-ph.IM]

\bibitem[\protect\citeauthoryear{{Ciecielag} \& {Chodorowski}}{{Ciecielag} \&
  {Chodorowski}}{2004}]{ps_ratio1}
{Ciecielag} P.,  {Chodorowski} M.~J.,  2004, \mnras, 349, 945

\bibitem[\protect\citeauthoryear{{Clifton}, {Ferreira}, {Padilla} \&
  {Skordis}}{{Clifton} et~al.}{2012}]{cfps2012}
{Clifton} T.,  {Ferreira} P.~G.,  {Padilla} A.,    {Skordis} C.,  2012,
  Phys.~Rept., 513, 1

\bibitem[\protect\citeauthoryear{{Colombi}, {Jaffe}, {Novikov} \&
  {Pichon}}{{Colombi} et~al.}{2009}]{powmes}
{Colombi} S.,  {Jaffe} A.,  {Novikov} D.,    {Pichon} C.,  2009, \mnras, 393,
  511

\bibitem[\protect\citeauthoryear{{Copeland}, {Sami} \& {Tsujikawa}}{{Copeland}
  et~al.}{2006}]{cst2006}
{Copeland} E.~J.,  {Sami} M.,    {Tsujikawa} S.,  2006, IJMPD, 15, 1753

\bibitem[\protect\citeauthoryear{{Davis}, {Li}, {Mota} \& {Winther}}{{Davis}
  et~al.}{2012}]{dlmw2012}
{Davis} A.-C.,  {Li} B.,  {Mota} D.~F.,    {Winther} H.~A.,  2012, \apj, 748,
  61

\bibitem[\protect\citeauthoryear{{Davis}, {Nusser}, {Masters}, {Springob},
  {Huchra} \& {Lemson}}{{Davis} et~al.}{2011}]{p_vel4}
{Davis} M.,  {Nusser} A.,  {Masters} K.~L.,  {Springob} C.,  {Huchra} J.~P.,
  {Lemson} G.,  2011, \mnras, 413, 2906

\bibitem[\protect\citeauthoryear{{de Felice} \& {Tsujikawa}}{{de Felice} \&
  {Tsujikawa}}{2010}]{dt2010}
{de Felice} A.,  {Tsujikawa} S.,  2010, Living Reviews in Relativity, 13, 3

\bibitem[\protect\citeauthoryear{{Deffayet}, {Esposito-Farese} \&
  {Vikman}}{{Deffayet} et~al.}{2009}]{dev2009}
{Deffayet} C.,  {Esposito-Farese} G.,    {Vikman} A.,  2009, \prd, 79, 084003

\bibitem[\protect\citeauthoryear{{Dvali}, {Gabadadze} \& {Porrati}}{{Dvali}
  et~al.}{2000}]{dgp2000}
{Dvali} G.,  {Gabadadze} G.,    {Porrati} M.,  2000, Phys.~Lett.~B, 485, 208

\bibitem[\protect\citeauthoryear{Faulkner, Tegmark, Bunn \& Mao}{Faulkner
  et~al.}{2007}]{ftbm2007}
Faulkner T.,  Tegmark M.,  Bunn E.~F.,    Mao Y.,  2007, Phys.Rev., D76, 063505

\bibitem[\protect\citeauthoryear{{Hellwing} \& {Juszkiewicz}}{{Hellwing} \&
  {Juszkiewicz}}{2009}]{rebel}
{Hellwing} W.~A.,  {Juszkiewicz} R.,  2009, Phys.~Rev.~D, 80, 083522

\bibitem[\protect\citeauthoryear{{Hellwing}, {Knollmann} \& {Knebe}}{{Hellwing}
  et~al.}{2010}]{rebel3}
{Hellwing} W.~A.,  {Knollmann} S.~R.,    {Knebe} A.,  2010, \mnras, 408, L104

\bibitem[\protect\citeauthoryear{{Hinterbichler} \& {Khoury}}{{Hinterbichler}
  \& {Khoury}}{2010}]{hk2010}
{Hinterbichler} K.,  {Khoury} J.,  2010, \prl, 104, 231301

\bibitem[\protect\citeauthoryear{{Hoyle}, {Schmidt}, {Heckel}, {Adelberger},
  {Gundlach}, {Kapner} \& {Swanson}}{{Hoyle} et~al.}{2001}]{gr_test2}
{Hoyle} C.~D.,  {Schmidt} U.,  {Heckel} B.~R.,  {Adelberger} E.~G.,  {Gundlach}
  J.~H.,  {Kapner} D.~J.,    {Swanson} H.~E.,  2001, Physical Review Letters,
  86, 1418

\bibitem[\protect\citeauthoryear{{Hu} \& {Sawicki}}{{Hu} \&
  {Sawicki}}{2007}]{hs2007}
{Hu} W.,  {Sawicki} I.,  2007, \prd, 76, 064004

\bibitem[\protect\citeauthoryear{{Hudson} \& {Turnbull}}{{Hudson} \&
  {Turnbull}}{2012}]{p_vel1}
{Hudson} M.~J.,  {Turnbull} S.~J.,  2012, \apjl, 751, L30

\bibitem[\protect\citeauthoryear{{Jain} \& {Zhang}}{{Jain} \&
  {Zhang}}{2008}]{jz2008}
{Jain} B.,  {Zhang} P.,  2008, \prd, 78, 063503

\bibitem[\protect\citeauthoryear{{Jennings}, {Baugh}, {Li}, {Zhao} \&
  {Koyama}}{{Jennings} et~al.}{2012}]{jblkz2012}
{Jennings} E.,  {Baugh} C.~M.,  {Li} B.,  {Zhao} G.-B.,    {Koyama} K.,  2012,
  ArXiv:1205.2698 [astro-ph.CO]

\bibitem[\protect\citeauthoryear{{Jennings}, {Baugh} \& {Pascoli}}{{Jennings}
  et~al.}{2011}]{ps_ratio2}
{Jennings} E.,  {Baugh} C.~M.,    {Pascoli} S.,  2011, \mnras, 410, 2081

\bibitem[\protect\citeauthoryear{{Jing}}{{Jing}}{2005}]{j2005}
{Jing} Y.,  2005, \apj, 620, 559

\bibitem[\protect\citeauthoryear{{Keselman}, {Nusser} \& {Peebles}}{{Keselman}
  et~al.}{2009}]{rebel2}
{Keselman} J.~A.,  {Nusser} A.,    {Peebles} P.~J.~E.,  2009, ArXiv:0912.4177
  [astro-ph.CO]

\bibitem[\protect\citeauthoryear{{Khoury} \& {Weltman}}{{Khoury} \&
  {Weltman}}{2004}]{kw2004}
{Khoury} J.,  {Weltman} A.,  2004, \prd, 69, 044026

\bibitem[\protect\citeauthoryear{{Kosowsky} \& {Bhattacharya}}{{Kosowsky} \&
  {Bhattacharya}}{2009}]{p_vel3}
{Kosowsky} A.,  {Bhattacharya} S.,  2009, \prd, 80, 062003

\bibitem[\protect\citeauthoryear{{Koyama}, {Taruya} \& {Hiramatsu}}{{Koyama}
  et~al.}{2009}]{kth2009}
{Koyama} K.,  {Taruya} A.,    {Hiramatsu} T.,  2009, \prd, 79, 123512

\bibitem[\protect\citeauthoryear{{Lee}, {Zhao}, {Li} \& {Koyama}}{{Lee}
  et~al.}{2012}]{lzlk2012}
{Lee} J.,  {Zhao} G.,  {Li} B.,    {Koyama} K.,  2012, arXiv:1204.6608
  [astro-ph.CO]

\bibitem[\protect\citeauthoryear{{Lewis}, {Challinor} \& {Lasenby}}{{Lewis}
  et~al.}{2000}]{camb}
{Lewis} A.,  {Challinor} A.,    {Lasenby} A.,  2000, \apj, 538, 473

\bibitem[\protect\citeauthoryear{{Li} \& {Barrow}}{{Li} \&
  {Barrow}}{2007}]{lb2007}
{Li} B.,  {Barrow} J.~D.,  2007, \prd, 75, 084010

\bibitem[\protect\citeauthoryear{{Li} \& {Barrow}}{{Li} \&
  {Barrow}}{2011}]{lb2011}
{Li} B.,  {Barrow} J.~D.,  2011, \prd, 83, 024007

\bibitem[\protect\citeauthoryear{{Li}, {Zhao}, {Teyssier} \& {Koyama}}{{Li}
  et~al.}{2012}]{lztk2012}
{Li} B.,  {Zhao} G.-B.,  {Teyssier} R.,    {Koyama} K.,  2012, \jcap, 1, 51

\bibitem[\protect\citeauthoryear{{Li} \& {Zhao}}{{Li} \& {Zhao}}{2009}]{lz2009}
{Li} B.,  {Zhao} H.,  2009, \prd, 80, 044027

\bibitem[\protect\citeauthoryear{{Li} \& {Zhao}}{{Li} \& {Zhao}}{2010}]{lz2010}
{Li} B.,  {Zhao} H.,  2010, \prd, 81, 104047

\bibitem[\protect\citeauthoryear{{Li} \& {Hu}}{{Li} \& {Hu}}{2011}]{lh2011}
{Li} Y.,  {Hu} W.,  2011, \prd, 84, 084033

\bibitem[\protect\citeauthoryear{{Lyne}, {Burgay}, {Kramer}, {Possenti},
  {Manchester}, {Camilo}, {McLaughlin}, {Lorimer}, {D'Amico}, {Joshi},
  {Reynolds} \& {Freire}}{{Lyne} et~al.}{2004}]{gr_test3}
{Lyne} A.~G.,  {Burgay} M.,  {Kramer} M.,  {Possenti} A.,  {Manchester} R.~N.,
  {Camilo} F.,  {McLaughlin} M.~A.,  {Lorimer} D.~R.,  {D'Amico} N.,  {Joshi}
  B.~C.,  {Reynolds} J.,    {Freire} P.~C.~C.,  2004, Science, 303, 1153

\bibitem[\protect\citeauthoryear{{Mota} \& {Shaw}}{{Mota} \&
  {Shaw}}{2007}]{ms2007}
{Mota} D.~F.,  {Shaw} D.~J.,  2007, \prd, 75, 063501

\bibitem[\protect\citeauthoryear{{Navarro} \& {Van Acoleyen}}{{Navarro} \& {Van
  Acoleyen}}{2007}]{nv2007}
{Navarro} I.,  {Van Acoleyen} K.,  2007, \jcap, 2, 22

\bibitem[\protect\citeauthoryear{{Nicolis}, {Rattazzi} \&
  {Trincherini}}{{Nicolis} et~al.}{2009}]{nrt2009}
{Nicolis} A.,  {Rattazzi} R.,    {Trincherini} E.,  2009, \prd, 79, 064036

\bibitem[\protect\citeauthoryear{{Nusser}, {Gubser} \& {Peebles}}{{Nusser}
  et~al.}{2005}]{NGP}
{Nusser} A.,  {Gubser} S.~S.,    {Peebles} P.~J.,  2005, \prd, 71, 083505

\bibitem[\protect\citeauthoryear{{Oyaizu}}{{Oyaizu}}{2008}]{oyaizu2008}
{Oyaizu} H.,  2008, \prd, 78, 123523

\bibitem[\protect\citeauthoryear{{Oyaizu}, {Lima} \& {Hu}}{{Oyaizu}
  et~al.}{2008}]{olh2008}
{Oyaizu} H.,  {Lima} M.,    {Hu} W.,  2008, \prd, 78, 123524

\bibitem[\protect\citeauthoryear{{Patiri}, {Betancort-Rijo} \&
  {Prada}}{{Patiri} et~al.}{2012}]{pbp2012}
{Patiri} S.~G.,  {Betancort-Rijo} J.,    {Prada} F.,  2012, \aap, 541, L4

\bibitem[\protect\citeauthoryear{{Peebles}}{{Peebles}}{1980}]{1980Peebles}
{Peebles} P.~J.~E.,  1980, {The large-scale structure of the universe}.
Research supported by the National Science Foundation.~Princeton, N.J.,
  Princeton University Press, 1980.~435 p.

\bibitem[\protect\citeauthoryear{{Perlmutter}, {Aldering}, {Deustua}, {Fabbro},
  {Goldhaber}, {Groom}, {Kim}, {Kim} \& {et~al.,}}{{Perlmutter}
  et~al.}{1999}]{petal}
{Perlmutter} S.,  {Aldering} G.,  {Deustua} S.,  {Fabbro} S.,  {Goldhaber} G.,
  {Groom} D.~E.,  {Kim} A.~G.,  {Kim} M.~Y.,    {et~al.,} 1999, \apj, 517, 565

\bibitem[\protect\citeauthoryear{{Pike} \& {Hudson}}{{Pike} \&
  {Hudson}}{2005}]{p_vel2}
{Pike} R.~W.,  {Hudson} M.~J.,  2005, \apj, 635, 11

\bibitem[\protect\citeauthoryear{{Pueblas} \& {Scoccimarro}}{{Pueblas} \&
  {Scoccimarro}}{2009}]{ps2009}
{Pueblas} S.,  {Scoccimarro} R.,  2009, \prd, 80, 043504

\bibitem[\protect\citeauthoryear{{Riess}, {Filippenko}, {Challis},
  {Clocchiattia}, {Diercks}, {Garnavich}, {Gilliland}, {Hogan} \&
  {et~al.,}}{{Riess} et~al.}{1998}]{retal}
{Riess} A.~G.,  {Filippenko} A.~V.,  {Challis} P.,  {Clocchiattia} A.,
  {Diercks} A.,  {Garnavich} P.~M.,  {Gilliland} R.~L.,  {Hogan} C.~J.,
  {et~al.,} 1998, \aj, 116, 1009

\bibitem[\protect\citeauthoryear{{Schaap} \& {van de Weygaert}}{{Schaap} \&
  {van de Weygaert}}{2000}]{sv2000}
{Schaap} W.~E.,  {van de Weygaert} R.,  2000, \aap, 363, L29

\bibitem[\protect\citeauthoryear{{Schmidt}}{{Schmidt}}{2009}]{schmidt2009}
{Schmidt} F.,  2009, \prd, 80, 043001

\bibitem[\protect\citeauthoryear{{Schmidt}, {Lima}, {Oyaizu} \& {Hu}}{{Schmidt}
  et~al.}{2009}]{sloh2009}
{Schmidt} F.,  {Lima} M.,  {Oyaizu} H.,    {Hu} W.,  2009, \prd, 79, 083518

\bibitem[\protect\citeauthoryear{{Schmidt}, {Vikhlinin} \& {Hu}}{{Schmidt}
  et~al.}{2009}]{svh2009}
{Schmidt} F.,  {Vikhlinin} A.,    {Hu} W.,  2009, \prd, 80, 083505

\bibitem[\protect\citeauthoryear{{Smith}, {Peacock}, {Jenkins}, {White},
  {Frenk}, {Pearce}, {Thomas}, {Efstathiou} \& {Couchman}}{{Smith}
  et~al.}{2003}]{halofit}
{Smith} R.~E.,  {Peacock} J.~A.,  {Jenkins} A.,  {White} S.~D.~M.,  {Frenk}
  C.~S.,  {Pearce} F.~R.,  {Thomas} P.~A.,  {Efstathiou} G.,    {Couchman}
  H.~M.~P.,  2003, \mnras, 341, 1311

\bibitem[\protect\citeauthoryear{{Sotiriou} \& {Faraoni}}{{Sotiriou} \&
  {Faraoni}}{2010}]{sf2010}
{Sotiriou} T.~P.,  {Faraoni} V.,  2010, Reviews of Modern Physics, 82, 451

\bibitem[\protect\citeauthoryear{{Springel}, {Yoshida} \& {White}}{{Springel}
  et~al.}{2001}]{Gadget1}
{Springel} V.,  {Yoshida} N.,    {White} S.~D.~M.,  2001, New Astronomy, 6, 79

\bibitem[\protect\citeauthoryear{Starobinsky}{Starobinsky}{2007}]{s2007}
Starobinsky A.~A.,  2007, JETP Lett., 86, 157

\bibitem[\protect\citeauthoryear{{Teyssier}}{{Teyssier}}{2002}]{ramses}
{Teyssier} R.,  2002, \aap, 385, 337

\bibitem[\protect\citeauthoryear{{van de Weygaert} \& {Schaap}}{{van de
  Weygaert} \& {Schaap}}{2009}]{vs2009}
{van de Weygaert} R.,  {Schaap} W.,  2009, in {Mart{\'{\i}}nez} V.~J.,  {Saar}
  E.,  {Mart{\'{\i}}nez-Gonz{\'a}lez} E.,   {Pons-Border{\'{\i}}a} M.-J.,  eds,
  Data Analysis in Cosmology Vol.~665 of Lecture Notes in Physics, Berlin
  Springer Verlag, {The Cosmic Web: Geometric Analysis}.
pp 291--413

\bibitem[\protect\citeauthoryear{{Wang}, {Caldwell}, {Ostriker} \&
  {Steinhardt}}{{Wang} et~al.}{2000}]{wcos2000}
{Wang} L.,  {Caldwell} R.~R.,  {Ostriker} J.~P.,    {Steinhardt} P.~J.,  2000,
  \apj, 530, 17

\bibitem[\protect\citeauthoryear{{Will}}{{Will}}{2006}]{w2006}
{Will} C.~M.,  2006, in {\it The Confrontation between General Relativity and
  Experiment} Vol.~9 of Living Rev. Relativity, {}.
p.~3

\bibitem[\protect\citeauthoryear{{Zel'dovich}}{{Zel'dovich}}{1970}]{za}
{Zel'dovich} Y.~B.,  1970, \aap, 5, 84

\bibitem[\protect\citeauthoryear{{Zhao}, {Li} \& {Koyama}}{{Zhao}
  et~al.}{2011}]{zlk2011}
{Zhao} G.-B.,  {Li} B.,    {Koyama} K.,  2011, \prd, 83, 044007

\end{thebibliography}

\bsp

%\onecolumn
%\appendix
%
%\section{Beyond the First-order Approximations}

\label{lastpage}

\end{document}